\documentclass[pdflatex, 12pt, sn-chicago]{sn-jnl}% Chicago-based Humanities Reference Style

\usepackage{appendix}
\usepackage{longtable} 
\usepackage[table]{xcolor} 
\usepackage{graphicx}%
\usepackage{multirow}%
\usepackage{amsmath,amssymb,amsfonts}%
\usepackage{amsthm}%
\usepackage{mathrsfs}%
\usepackage{xcolor}%
\usepackage{textcomp}%
\usepackage{manyfoot}%
\usepackage{booktabs}%
\usepackage{tikz}
\usepackage{algorithm}%
\usepackage{algorithmicx}%
\usepackage{amsmath}
\usepackage{algpseudocode}%
\usepackage{listings}%
\usepackage{subcaption}
\usepackage{amsmath}
\usepackage{amssymb}
\usepackage{algorithm}

\usepackage{algpseudocode}
\usepackage[utf8]{inputenc}
\usepackage{amssymb, amsmath}
\usepackage{graphicx}   
\usepackage{booktabs}   
\usepackage{array}

\usepackage[ruled,vlined,algo2e]{algorithm2e}
\SetKwInOut{Output}{Output}
\SetKwInOut{Data}{Data}

\usepackage{tikz}
\definecolor{grey5}{rgb}{0.9,0.9,0.9}
\definecolor{grey4}{rgb}{0.833,0.833,0.833}
\definecolor{grey3}{rgb}{0.766,0.766,0.766}
\definecolor{grey2}{rgb}{0.7,0.7,0.7}
\definecolor{grey1}{rgb}{0.633,0.633,0.633}
\definecolor{darkgrey}{rgb}{0.3,0.3,0.3}

\usetikzlibrary{backgrounds}
\usetikzlibrary{calc}

\usepackage{hyperref}
\usepackage{pdfpages}
\theoremstyle{thmstyleone}%

\theoremstyle{thmstyletwo}%

\theoremstyle{thmstylethree}%

\begin{document}

\title[Tripartite social networks]{Modeling Tripartite Hyperevents in Scientific Collaboration Networks}
\author*[1]{\fnm{Amin Gino} \sur{Fabbrucci Barbagli}}\email{amingino.fabbruccibarbagli@phd.units.it}

\author[2]{\fnm{Jürgen} \sur{Lerner}}

\author[3]{\fnm{Viviana} \sur{Amati}}

\author[1]{\fnm{Domenico} \sur{De Stefano}}

\affil*[1]{\orgdiv{Department of Social and Political Science}, \orgname{University of Trieste}, {\country{Italy}}}

\affil[2]{\orgdiv{Department of Computer Science}, \orgname{University of Konstanz}, \orgaddress{\country{Germany}}}

\affil[3]{\orgdiv{Department of Statistics and Quantitative Methods}, \orgname{University of Milano-Bicocca}, \orgaddress{\country{Italy}}}

\abstract{Sociological research has framed collective action in science, innovation, and culture as tripartite networks connecting teams of actors, lists of prior works, and sets of labels (e.g., keywords, topics). While methods for multipartite social networks were proposed decades ago, and have received a recent surge in interest, none of the suggested solutions scale to the size and granularity of contemporary data sets (scientific publications, patents, filmmaking) and at the same time allow for testing multiple competing hypotheses about the drivers of collective production. In this paper, we address this gap by applying Relational Hyperevent Models (RHEM) to dynamic tripartite hypergraphs. Using scientific networks as a case study, we model events linking any number of actors, references, and keywords, testing and controlling for inter-dependencies within and between each set.} 

\maketitle
\section{Introduction}
Scientific knowledge and production arise from interactions among researchers, between prior knowledge and innovation, and between context and specific domain \citep{latour1987science, fortunato2018science}. A scientific outcome is the result of collaboration among researchers who generate content that builds on prior knowledge, which identifies the topic and contribution of their work using keywords. 

One way to represent and describe the emergence of scientific knowledge is to use concepts and tools from network analysis \citep{wasserman1994social,borgatti2009network}. Over the past decades, it has been used to analyze interactions among different types of entities in the domain of science and innovation (for a review, see \cite{kang2023scientific} and \cite{bellotti2025scientific}, and the reference therein). Its application has been boosted over the last twenty years by the increasing availability of metadata on scientific publications and discoveries.

Network analysis has been applied to scientific collaboration networks to explain how entities such as researchers and institutions interact to produce content. Much of the work on scientific collaboration networks focuses on formal interactions (tangible signs of interactions which are considered proxies for collaboration) such as co-authorship or joint collaboration in research projects. %Little has been  done in terms of informal interactions (connections that do not lead to a scientific product per se, such as advice, discussions, and supervision) \citep{crane1977social}. 
In this scenario, co-authorship networks (nodes are researchers and a link between two actors exists if they have co-authored at least one paper) have been analyzed to describe the network structure among scholars \citep{lu2009measure,abbasi2012egocentric,de2013use}, to reveal the presence of teams or communities \citep{ghosh2017examining,cugmas2019scientific,geremia2025community} and to determine the mechanisms explaining the formation of the collaboration and its evolution over time in different domains \citep{mali2011dynamic,fagan2018assessing,espinosa2025socio,fabbrucci2025unveiling}. 
Co-authorship networks or other types of formal interactions among researchers are not the only ways to explore scientific production and its evolution. Another application of network analysis in the scientific production domain concerns citation networks where nodes are scientific publications and relationships are references to other publications \citep{price1965networks, radicchi2011citation,filippi2023drivers}. The analysis of these networks investigates how knowledge evolves over time by building on prior work and identifies key publications, innovations and discoveries. 

Finally, in scientific collaboration and production, network analysis has also been adopted to identify emerging topics from the labels (e.g., tokens extracted from titles or abstracts through appropriate textual data processing)  or keywords researchers use to characterize their works \citep{lozano2019complex,schafermeier2023research}. In this context, nodes may be represented by keywords and their relationships are determined by their co-occurrence in the same publication or in different publications.

Collective scientific production, therefore, can be explored in terms of co-authorship, citation, or keywords co-occurrence networks. These structures are usually represented as one-mode networks characterized by a single set of nodes of the same type and the relationships among them.

\subsection{From one mode network to hypergraph}

The analysis of one-mode networks in this context faces two important limitations:
the first concerns the level of analysis. By examining pairs of researchers, references and keywords, standard approaches overlook the the inherently multicast nature of such interactions and the fact that the observed ties may be originated from just one scientific product instead of implying more complex interactions, such transitivity or closure (e.g., in a co-authorship network a couple of authors whose connections depend on the fact that they shared a third common author in a past publication). For instance, a scientific product generated by researchers (e.g., A, B and C), necessarily determine links between each all dyads (AB, AC, BC), however this cannot be considered as a transitive triad \citep{kim2017over}. A similar issue affects ties if we consider networks of cited papers and keywords co-occurrence. 

To address these issues, some scholars may use bipartite networks (networks in which there are two distinct sets of nodes and only relationships from one set to another are allowed) with links between researchers, keywords, references and scientific products. However, the application of bipartite networks is only a convenient representation because subsequent analyses are usually performed by projecting the two-mode networks onto one-mode networks \citep{opsahl2013triadic}, as testified not only in networks emerging from scientific publications but also by the collaboration in patents \citep{capellari2016academic}, in funded research project interactions \citep{Morea_2024} or networks in cultural productions such as movies industry \citep{burgdorf2024communities}. This approach further extends to multi-mode networks where three or more distinct types of entities and their interactions are considered, even though the projection inherently entails information loss and precludes the examination of competing relational mechanisms simultaneously \citep{everett2013dual}. 

The second limitation lies in the fact that the co-authorship network, the citation network, and the keyword network are typically analyzed separately, thus neglecting their interdependencies.
Researchers become recognizable through their use of keywords and citation patterns; keywords take their meaning from the documents and scholars related to them, and publications gain importance from their conceptual framing and social embeddedness. Folksonomy \citep{hotho2006folksonomy} exemplifies such interdependence: the users' activity and tagging behavior lead to a tripartite structure of actors, descriptors, and artefacts \citep{lambiotte2006collaborative, trant2009folksonomy, hotho2006folksonomy}. This conceptualization fits into the broader framework of Breiger's theory of duality \citep{breiger1974duality} and its generalization \citep{lee2018doorway} illustrating how actors and collectives mutually constitute one another -- e.g., individuals are defined by groups they belong to, and groups are defined by their members. The disjoint analysis provides only a partial view and misses the inherent interdependencies that characterize scientific knowledge production.

Given these issues, the analysis of the dynamics of collective scientific production based on longitudinal observations of one-mode networks remains limited, as one-mode networks cannot properly capture the evolution of co-authorship, citations, and keywords co-occurrence.

To overcome these issues, researchers have recently used hypergraphs, which are networks connecting multiple entities of the same type and where every connection in the subset of nodes constitutes a hyperedge. Recently, \citet{ko2022growth} analyzed co-authorship networks to describe general features of ``real-world'' hypergraphs. \citet{lerner2019remdyadsrelationalhyperevent}, \citet{lerner_lomi2023}, \citet{fabbrucci2025unveiling} and \citet{boschi2025linearitytimehomogeneityrelationalhyper} used Relational Hypervent Models (RHEM) to investigate the determinants of co-authorship dynamics. \citet{espinosa2025socio} and \citet{lerner2025relational} recently extended this work to the joint analysis of co-authorship and citations. \citet{shi2015weaving} investigated interactions among different types of entities (authors, chemicals, diseases, and methods) in biomedicine using descriptive statistics and the random walk model to assess the evolution of biomedicine science. 

\begin{figure}
    \centering
    \includegraphics[width=0.8\linewidth]{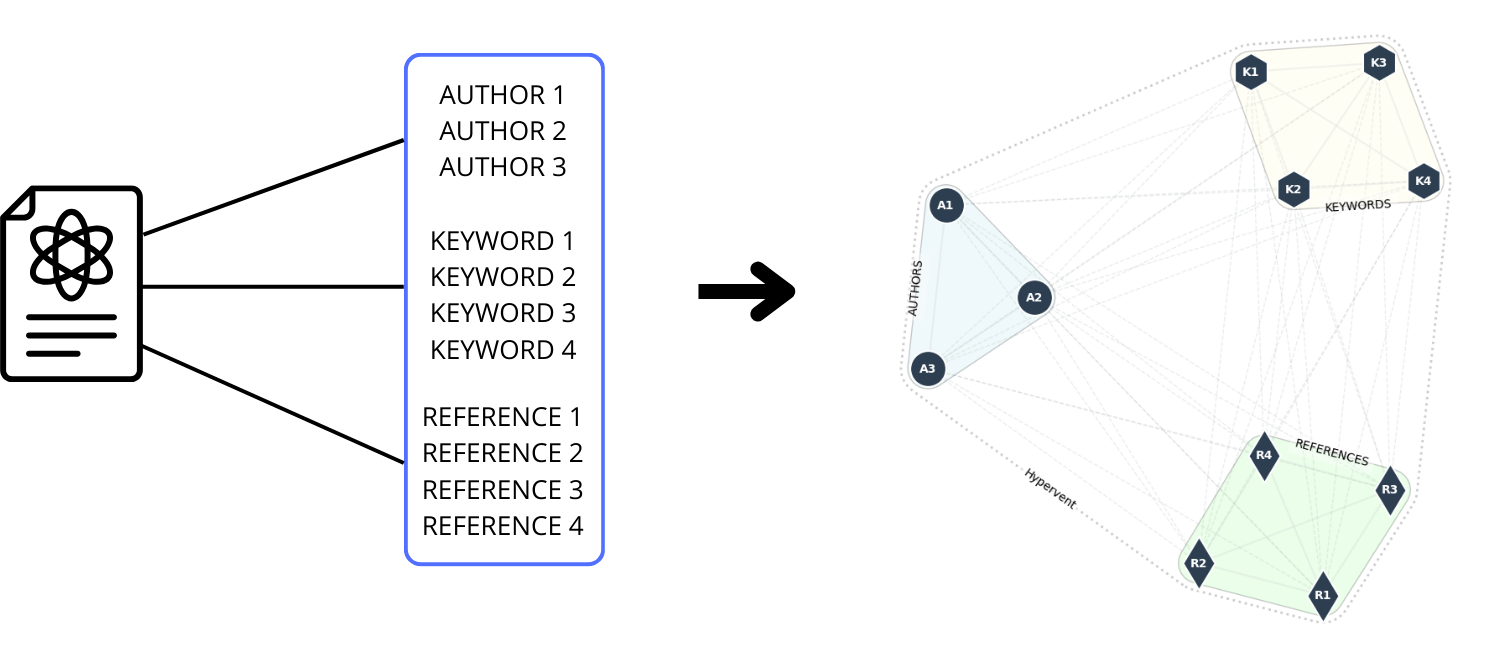}
    \caption{A hyperevent is a higher-order interaction that involves multiple entities of the same or different types and the relation that exists between them. In our case, the hyperevent is composed of three different hyperedges with the following relations: (1) authors and their references; (2) author and their keywords; (3) the relation between the authors' keywords and their references; 4)the relations within each set. }
    \label{fig:intro}
\end{figure}

Building on this latter stream of literature, in the present paper, we aim to consider the complex dynamics embedded in scientific collective production.
This poses several theoretical and methodological challenges. 
%, and we therefore propose an extension of the classical specification of RHEM.
Following the work of \citep{shi2015weaving}, we conceptualize a paper as a hyperevent connecting distinct sets of nodes: authors, references, and keywords (See Figure \ref{fig:intro}). Hereafter, we refer to this structure as ``tripartite hyperevent'' although in formal graph theory hypergraphs are defined on a single set of entities in our case it is formed by three different entities types. Node subsets--namely authors, references and keywords--form hyperevents that capture their interactions, both within and between the same subset of nodes (collaboration, co-citation, and keyword co-occurrence). For instance, a subset comprising academics and their references, along with co-authorship and citation relationships, might indicate whether co-authors share the same references and exhibit homophilous citation patterns.  

The resulting scientific network can be considered an extension of dynamic folksonomy, where sets of authors, papers, and keywords are linked between and within them. Authors are linked to co-authors, keywords and references of their published papers over time \citep{lerner2025relational, lerner2019remdyadsrelationalhyperevent, boschi2025linearitytimehomogeneityrelationalhyper}.

Since such a tripartite hypervent structure in the scientific network setting is intrinsically dynamic, we explore the co-evolution of relationships within and between the three types of entities involved by properly extending RHEM specifications and related effects \citep{lerner2019remdyadsrelationalhyperevent}.

This extension allows us to overcome the issues determined by using the one-mode or even two-mode settings, and the issues mentioned above. RHEMs provide a rigorous framework for testing hypotheses about the mechanisms underlying the occurrence of hyperevents, given time-stamped hyperedges that connect multiple entities simultaneously and dynamically. They are also capable of addressing issues of scalability related to the size and granularity of data (e.g., scientific publications, patents, filmmaking collaborations). Furthermore, with this family of statistical models, we aim to study not only the evolution of each entity but also the co-evolution of the three entities. This application extends to all cases in which dynamic interdependencies exist among different actors and domains. By analyzing these co-evolutionary patterns, we can identify how local mechanisms aggregate to shape the global emerging structure of the entire network.

From a structural standpoint, the co-authorship network is both observable and symmetric, whereas citation networks are observable but asymmetric. By contrast, co-occurrence networks—such as keyword associations—are theoretically inferred rather than directly observed.

The paper is organized as follows. Section 2 introduces the extended RHEM for tripartite hyperevents. In Section 3, we describe the case study and the related research question. Finally, in Section 4, we present the results and the final discussions.

\section{Extending Relational Hyperevent Model for tripartite hyperevent dynamics}

While methods for multipartite social networks have been proposed decades ago, and received a recent surge in interest, none of the suggested solutions scale to the size and granularity of data sets (scientific publication, patents, filmmaking) and at the same time allow for testing multiple competing hypotheses about the drivers of collective production.
%Tripartite structure presents challenges that approaches have not fully addressed. This 
Tripartite structures require a framework capable of representing higher-order interactions and dependencies among heterogeneous entities.
Relational hyperevent models (RHEMs), a recent family of statistical models defined within a point-process framework for relational event data \citep{perry2013point}, are suitable models to address these challenges. It can handle large-scale, fine-grained time-stamped data and assesses the propensity of actors to continue their interactions over time, while accounting for different nodes' characteristics \citep{lerner2019remdyadsrelationalhyperevent} and is suitable for polyadic settings and large networks \citep{lerner_lomi2023}.

In this study, we examine publication events that connect nodes of three types, specifically authors, references, and keywords, through distinct types of relations and the evolution of their interactions.

Unlike previous studies regarding hypergraph in scientific collaboration studies \citep{hancean_Lerner,espinosa2025socio}, we aim at modeling the coevolution of these relationships, which simultaneously helps explain scientific behaviors through three distinct but interrelated processes: team formation in co-authorship networks, knowledge building in co-citation networks and topic evolution in keywords co-occurrence networks. These networks, when analyzed jointly, provide information on scientific orientation, impact, and intellectual inheritance. Conceptually, this extension enables multiple interpretative frameworks: authors being recognizable through their keywords and the references they cite.
Keywords acquire meaning from the cited literature and the authors who adopt it. Paper visibility is driven by keywords and topic embedding. 

Formally, this framework can be represented as a hypergraph with different types of nodes, that are the three sets representing the key elements of scientific production embedded in a publication.

 Hypergraphs are a generalization of graphs where each hyperedge can join any number of vertices. A hypergraph $H$ is formally defined as an ordered pair of $H=(V,E)$ where: 
\begin{itemize}
    \item $V={v_{1}, v_{2}, v_{3}\dots, v_{n}}$ represents a set of nodes
    \item $E={e_{1}, e_{2}, e_{3}\dots, v_{n}}$ is a no-empyt substef of $V$, representing the set of hyperdges.
\end{itemize}
In the tripartite hypergraph, the vertex set $V$ is partitioned into three disjoint sets, that is 
$V=I_{} \cup J_{} \cup W_{}$:
\\
with
\begin{itemize}
    \item $I={i_{1}, i_{2}, i_{3}, \dots, i_{n}}$ as sets of authors;
    \item $J={j_{1}, j_{2}, j_{3}, \dots, j_{n}}$ as set of papers;
    \item $W={w_{1}, w_{2}, w_{3}, \dots, w_{n}}$ as set of keywords.
\end{itemize}

Each $e \in E$ is a single paper, resulting in a subset of $V$ that aggregates multiple elements of each entity.
\begin{equation*}
V=I_{e} \cup J_{e} \cup W_{e}
\end{equation*}
\begin{itemize}
    \item $I_{e} \subseteq I$ is a non-empty set of authors
    \item $J_{e} \subseteq J$ is the set of cited references 
    \item $W_{e} \subseteq W$ is the set of keywords associated to the paper.
\end{itemize}

The resulting hypergraph captures relations between these sets and within them. 

It can be modeled as a hypergraph evolving over time by adding a time $t$ corresponding to the year of publication. Finally, the hypergraph $H(t)=(I, J, W, t)$, results in a sequence of polyadic relational events that can be modeled using the Relational Hyperevent models (RHEM) \citep{lerner2019remdyadsrelationalhyperevent}.

As demonstrated in previous research, the model is adaptable to polyadic settings and large-scale networks \citep{lerner_lomi2023}. In particular, RHEMs can effectively handle fine-grained time-stamped events, such as those found in co-authorship networks comprising papers published across different years \citep{hancean_Lerner}.

Let $\mathcal{I}$ be a finite set of authors, $\mathcal{J}$ be a finite set of cited papers, and $\mathcal{W}$ be a finite set of keywords. For a point in time $t > 0$, let $\mathcal{I}_{t} \subseteq \mathcal{I}$ denote the set of authors who could potentially publish a paper at time $t$, and let $\mathcal{J}_{t} \subseteq \mathcal{J}$ denote the set of cited papers that could potentially be cited by a paper published at time $t$. Finally, let $\mathcal{W}_{t} \subseteq \mathcal{W}$ denote the set of keywords that could potentially be used in a paper published at time $t$.

The sequence of relational hyperevents is defined as:
\begin{equation*}
    j_m= (t_{1}, I_{1}, J_{1}, W_{1}),\ldots,(t_{n}, I_{n}, J_{n}, W_{n})
\end{equation*}
where $(t_m, I_m, J_m, W_m)$ indicates that authors $I_m \subseteq \mathcal{I}_{t_m}$ publish paper $j_m \in \mathcal{J}$ at time $t_m$, citing papers $J_m \subseteq \mathcal{J}_{t_m}$ and using keywords $W_m \subseteq \mathcal{W}_{t_m}$. %[JL alternative: write $j_m=(t_m,I_m,J_m,W_m)$ to indicate that the published paper \textbf{IS} the event.]

The event rate (also referred to as \textit{hazard rate} or \textit{intensity}) \citep{lerner2019remdyadsrelationalhyperevent} for hyperedge $e$ at time $t$, conditional on the network of past events, is specified as follows:

\begin{equation}
   \lambda(t,e,\beta,\mathcal{G}[E;t]) 
   = \lambda_{0}(t) \cdot \lambda_{1}(t,e,\beta,\mathcal{G}[E;t])
\end{equation}

\begin{equation}
   \lambda_{1}(t,e,\beta,\mathcal{G}[E;t]) 
   = \exp\!\left(\sum_{i=1}^{k} \beta_{i} \cdot x_{i}(t,e,\mathcal{G}[E;t])\right)
\end{equation}

\noindent where:
\begin{itemize}
    \item $\lambda_{0}(t)$ is the \textit{baseline hazard}, common to all hyperedges $e$, typically left unspecified;
    \item $\lambda_{1}(t,e,\beta,\mathcal{G}[E;t])$ is the \textit{relative hazard}, which determines the proportional probability that an event at time $t$ occurs on hyperedge $e$;
    \item $x_{i}(t,e,\mathcal{G}[E;t])$ are the covariates  describing how hyperedge $e$ is embedded in the network of past events $\mathcal{G}[E;t]$;
    \item $\beta = (\beta_{1},\dots,\beta_{k})$ is the parameter vector, where $\beta_{i} > 0$ indicates that the corresponding covariate increases the relative hazard, and $\beta_{i} < 0$ indicates a decrease.
\end{itemize}

Based on the observed event sequence $E$, the partial likelihood is given by:

\begin{equation}
L(\beta) = \prod_{m=1}^{n}\frac{\exp\bigl(\beta^\top x_{t_m}(I_m, J_m, W_m)\bigr)}{\displaystyle\sum_{(I, J, W) \in \binom{\mathcal{I}_{t_m}}{|I_m|} \times \binom{\mathcal{J}_{t_m}}{|J_m|} \times \binom{\mathcal{W}_{t_m}}{|W_m|}} \exp\bigl(\beta^\top x_{t_m}(I, J, W)\bigr)}
\end{equation}

\noindent where $\binom{\mathcal{I}_{t_m}}{|I_m|}$ denotes the set of all subsets of $\mathcal{I}_{t_m}$ of size $|I_m|$, and analogously for $\mathcal{J}_{t_m}$ and $\mathcal{W}_{t_m}$.

Given the values of the statistics $x_{i}(t,e,\mathcal{G}[E;t])$ for all elements of the risk sets $R_{t_e}$ at the event times $t_{e}$, maximum likelihood estimates can be computed using standard statistical software \citep{lerner2019remdyadsrelationalhyperevent}.

%%%%%%%%%%%%%%%%%%%%%%%%%%%
\subsection{Closure}

\begin{figure}
    \centering
    \includegraphics[width=1\linewidth]{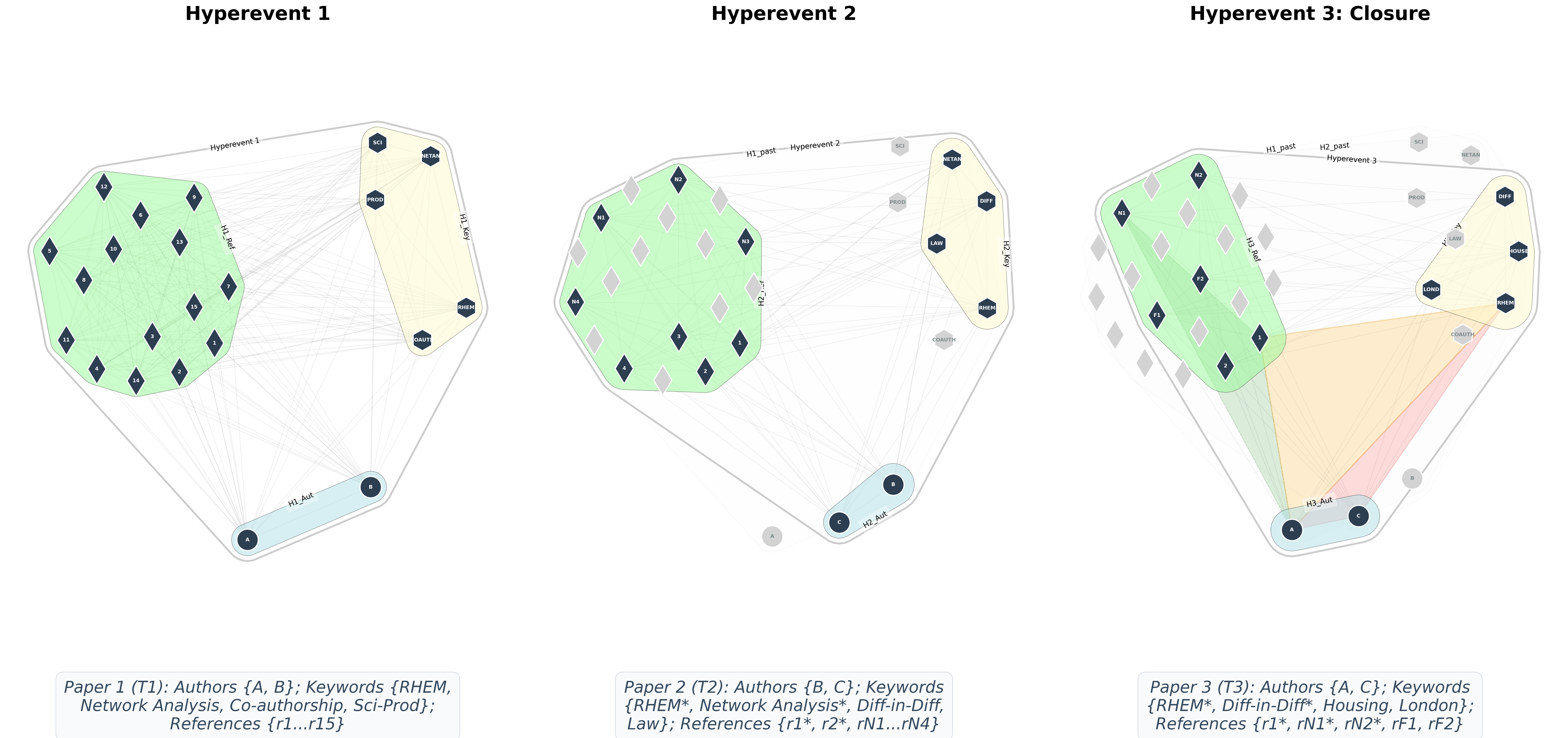}
\caption{The figure illustrates a sequence of three relational hyperevents ($T_1, T_2, T_3$), where nodes represent authors (circles), keywords (hexagons), and references (diamonds). Hyperevent 1 represents the initial collaboration co-authored by authors $A$ and $B$, with keywords RHEM, COAUTH, and NETAN and references $r_1$ to $r_{10}$. At time $T_2$, authors $B$ and $C$ initiate a collaboration, where some of the keywords and citations from the previous hyperevent were used and new entities were added from author $C$'s domain, thus generating a partial structural overlap. At time $T_3$, authors $A$ and $C$ publish a paper, where selected keywords and references from both $T_1$ and $T_2$ (denoted by an asterisk) alongside novel entities have been used. 
In time $T_3$, closure is observed highlighted with a red area for the closure auth.auth.keywords, green for the auth.ref.ref and orange for the aut.ref.key. A summary of all the closures considered is reported in Tab.1.}
    \label{fig:closure}
\end{figure}

Closure is a fundamental statistic in RHEM that indicates the transition from an open to a denser, more connected structure, in which past events increase the likelihood that entities will be connected in future events. It allows us to study the co-evolution of the hyperevent structure by testing the transformation of the different entity types within and between them.
In this work, we propose different combinations of closure, including tripartite closure, in which three distinct types of entities participate jointly in polyadic hyperevents, as well as closure within distinct hyperedges. 
In general, closure occurs when two nodes that are not directly connected interact via a shared actor. 
Following the definition of closure proposed in prior work \citep{hancean_Lerner, lerner2019remdyadsrelationalhyperevent, espinosa2025socio}, we consider a two-path configuration $v_1$--$v_2$--$v_3$, in which the intermediary node $v_2$ connects $v_1$ and $v_3$ indirectly. Closure occurs when the endpoint pair $(v_1, v_3)$ subsequently co-participates in a hyperevent, thereby completing a closed triad.

In tripartite hyperevents, nodes may belong to one of the three distinct types or to only one. We adopt a two-level classification:
the outer level specifies 
the types of the endpoint nodes $v_1$ and $v_3$, while the inner level 
specifies the type of the intermediary $v_2$. This yields a systematic enumeration 
of closure effects, each corresponding to a distinct relational mechanism in the 
co-evolution of scientific networks. %not sure if we want to include it

For example, if author A has collaborated with author B in one publication at time $t_1$, and author B has collaborated with author C in another work at time $t_2$, if at time $t_3$ A and C start a collaboration, closure will assume a positive value. These statistics can be used to show the social mechanisms underlying scientific collaboration: shared collaborators facilitate new collaborations, build trust, and provide opportunities for joint work.
In RHEM, the core idea is that each publication is not isolated; it depends on the network's history and suggests particular network behavior. Past events are a fundamental element in the computation of these statistics.

\begin{figure}
    \centering
    \includegraphics[width=0.8\linewidth]{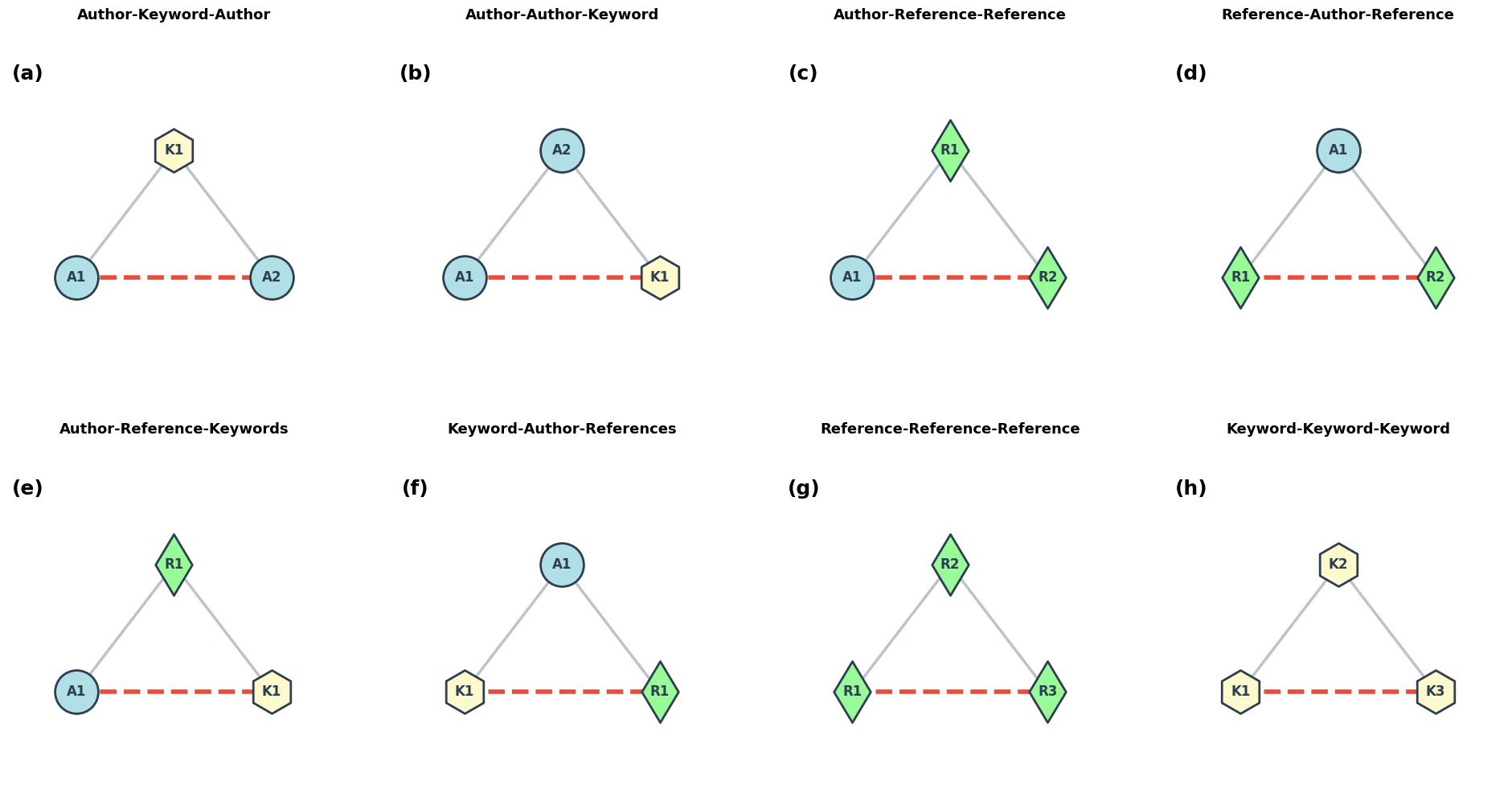}
    \caption{Examples of triadic closure effects involving one entity or multiple entities. In general, we have as a precondition a two-path of
three nodes $v1$–$v2$–$v3$ and then the pair $(v1, v3)$ participates in the same event, that is, closes the two paths.
The “outer” classification distinguishes the types of the two outer nodes $v1$ and $v3$ and then the “inner” classification distinguishes some possibilities for the third node.}
    \label{fig:closure_explanation}
\end{figure}

For example, “closure.aut.key.ref” refers to a two path $i–k–j$ and the closure effect means that an author $i$ cites a paper $j$
because previously $i$ published a paper labeled by keyword $k$ and $k$ labeled a paper that cited $j$ (a complete list is reported in Table \ref{tab:effects}). Under this assumption, we observe, e.g., the closure \textit{Author-Keyword-Author} when authors tend to adopt keywords used by their collaborators.
As illustrated in the Figure \ref{fig:closure} (red area) and Figure \ref{fig:closure_explanation} a this occurs when two authors ($A$ and $C$) who have independently used shared common keyword in the past ($T_1$ and $T_2$), publish a paper together at time $T_3$ using the same keywords, e.g. (${RHEM}$), producing a semantical closure. In other words, by adopting someone's keywords, semantic proximity (or shared expertise/topic interests) is introduced, and a positive closure is translated into topic or research area proximity. Hyperedges begin to overlap (and become denser) and increase the propensity for persistent keyword sharing, since they already operate within the same methodological area. 

Differently, if we look for closure \textit{Author-Author-Keyword} (see Figure \ref{fig:closure_explanation}b), author $A$ and $C$ are indirectly connected to a keyword $K$ via a common collaborator. In this case, the closure is positive if author $A$ uses a keyword that was previously used by a co-author in a joint work with him/her. This process indicates a knowledge sharing, in particular, of shared methodological application or technical language.

The closure \textit{Author-Reference-Reference} (see Figure \ref{fig:closure_explanation}c) occurs when a reference acts as an intermediary node connecting an author ($A$) and a reference that will be jointly cited by the author in a new event ($T_3$). This might suggest, for example, that authors build on existing knowledge by integrating prior citations into new research.

Closure \textit{Reference-Author-Reference} (see Figure \ref{fig:closure} green area and Figure
\ref{fig:closure_explanation}d) occurs when an author acts as an intermediary node bridging together two references, previously present disjointly in past events, by citing them in a new event. The authors build on past knowledge and unify it in a unique framework
by defining a topic evolution in a more stable knowledge paradigm, facilitating its consolidation.

Closure \textit{Author-Reference-Keywords} (See Figure \ref{fig:closure} orange area and Figure \ref{fig:closure_explanation}e) occurs when a reference ($r_1$) is the intermediary node connecting an author $A$, and a keyword ($k_1$), used in the past events that are now jointly used in a subsequent event. In other words, the author $A$ has previously cited $r_1$, which was already associated with the keyword $k_1$, and uses it jointly in the new event.
This mechanism can represent the formalization of expertise in a specific domain based on the related literature. By integrating the related bibliography into the corresponding semantic vocabulary, the authors achieve a topic consolidation of the research, connecting bibliographic references to keywords. 
On the other hand, the tripartite closure \textit{Keyword-Author-References} (see Figure \ref{fig:closure_explanation}f) occurs when an author $A$ is the intermediary node between a keyword and a reference that have previously appeared in past events and appear together in a future event. Thus, authors who have used a keyword $k_1$ in the past and a reference $r_1$ in the past, combine them into a new hyperevent.
Authors connect their research topic by a stabilized set of keywords and references, resulting in a more specialized research path.

Other closures occur within the same node type. Closure \textit{Reference-Reference-Reference} (see Figure \ref{fig:closure_explanation}g) suggests that documents which share references are more likely to be co-cited. Author A and Author B use the same references in $T_1$, and Author B and Author C use them in $T_2$; a latent knowledge cluster is formed. At $T_3$, we observe that a partial set of that knowledge is then reused. This suggests the existence of a shared background, where the positive closure stands for a bibliography denser hyperevent that could be translated into an ensemble of citations that need to be cited because they represent the foundational paradigm of that knowledge. Thus, publications sharing a common bibliographic foundation are more likely to be co-cited in subsequent events.

Closure \textit{Keyword-Keyword-Keyword} (see Figure \ref{fig:closure_explanation}h) represents the occurrence (or not) of keywords that have been jointly used in the past and appear to be used together. This represents the first element of the consolidation of the knowledge paradigm, in which elements tend to co-occur, forming a consolidated set of knowledge.

In general, a positive closure effect indicates that pairs of nodes that share common intermediaries are more likely to co-occur in future hyperevents. In contrast, a negative effect suggests the opposite tendency.

Closure indicates the propensity for pairs of nodes that share common intermediaries to increase or decrease the likelihood of co-occurring in subsequent hyperevents. To study the evolution of tripartite structure closure has been specified with different inter-type specification by focusing on reference, keywords or authors allowing to test evolving structure that emerge in the tripartite structure, and whether nodes involved in similar patterns of multi-type interactions are more likely to participate jointly in future events, or whether regularities observed within one mode influence the formation of ties across the others. 
This form of cross-type closure enables us to address polyadic relational events, overcoming the limitations of dyadic relational structures, which are constrained to pairwise interactions and two-mode settings. Different combinations of closure have been made to test and study the co-evolution of the tripartite structure (See Table 1).

\subsection{Subset Repetition with Geometrically Weighted Subset Repetition (GWSR)}
In this work, we also introduce the Geometrically Weighted Subset Repetition (GWSR) \citep{fabbrucci2025co}. Building on the definition of Subset repetition \citep{lerner2019remdyadsrelationalhyperevent} and the geometrically-weighted degree and geometrically-weighted edgewise shared partner statistics (GWESP) \citep{Hunter2007,Snijders2006,Levy2016,Handcock2008} to solve issues related to the standard subset repetition as: 1) the increasing weight of the binomial coefficient $\binom{|I \cap I_m|}{p}$ to weight past events; 2) the multicollinearity arising from the nested structure of the different order of the parameters $p$; 3) the lack of scaling across different hyperedges sizes. To overcome these issues, the GWSR was introduced.
\\

  Let a sequence of relational hyperevents be given by:
\[
E=(t_1,I_1,J_1),\dots,(t_n,I_n,J_n)\enspace,
\]
where $t_m$ is the time of the $m$-th hyperevent, $I_m \subseteq \mathcal{I}_{t_m}$ are the participating source nodes, and $J_m \subseteq \mathcal{J}_{t_m}$ are the participating target nodes.

Given a time $t$ and a set of nodes $(I, J)$, a subset repetition of order $(p, q)$ occurs when a set of sources $p$ repeatedly sends events to a set of targets $q$. It is defined as:
\[
sub.rep^{(p,q)}(t,I,J) = \frac{1}{\binom{|I|}{p}\binom{|J|}{q}} \cdot \sum_{I' \in \binom{I}{p}} \sum_{J' \in \binom{J}{q}} hy.deg(t,I',J')\enspace,
\]
where the ``hyperedge degree'' (\textit{hy.deg}), ignoring any temporal decay, is defined by:
\[
hy.deg(t,I',J') = \sum_{t_m < t} \mathbf{1}(I' \subseteq I_m \land J' \subseteq J_m)\enspace.
\]
In other words, the hyperedge degree quantifies the frequency of prior events $(t_m, I_m, J_m)$ satisfying the condition $I' \subseteq I_m$ and $J' \subseteq J_m$—representing the instances in which all nodes in the subsets $I'$ and $J'$ jointly participated.

The sets $\mathcal{I}_t$ and $\mathcal{J}_t$ may but do not have to be disjoint. An example of a two-mode hyperevent is the publication of scientific papers, where $I_m$ denotes the set of authors of the paper and $J_m$ denotes the set of its references. In this case, the sources and targets are selected from disjoint sets. While the sender $i_m$ is typically not among the receivers $J_m$ (``loopless directed hyperevents''), they are typically selected from the same set of ``actors''. This implies that past senders may become future receivers, and vice versa, so that in the case of directed one-mode hyperevents, effects such as ``(subset) reciprocation'' are possible. In contrast, reciprocation or subset reciprocation are not possible for two-mode hyperevents.

A key property of subset repetition is that the contribution of each past hyperevent $(I_m, J_m)$ to $sub.rep^{(p,q)}(t,I,J)$ is determined by the magnitude of its overlaps with $I$ and $J$. Specifically, the weight of a past event equals $\binom{|I \cap I_m|}{p}\binom{|J \cap J_m|}{q}$, which counts the number of subset pairs of size $(p,q)$ shared by both the current and the past event. Following the standard convention that $\binom{k}{p} = 0$ for $k < p$, hyperevents with overlaps smaller than $p$ in the source set or $q$ in the target set do not contribute to the statistic.

This leads to an equivalent formulation of subset repetition of order $(p,q)$:
\begin{equation}
  \label{eq:subrep_pq}
  sub.rep^{(p,q)}(t,I,J) = \frac{1}{\binom{|I|}{p}\binom{|J|}{q}} \cdot \sum_{t_m < t} \binom{|I \cap I_m|}{p} \binom{|J \cap J_m|}{q}.
\end{equation}

Equation~\ref{eq:subrep_pq} highlights a fundamental limitation of subset repetition statistics that the GWSR is designed to overcome. While it is intuitive that past hyperevents $(I_m, J_m)$ with larger overlaps with the focal hyperevent $(I,J)$ should be assigned greater weight, the product of binomial coefficients $\binom{|I \cap I_m|}{p} \binom{|J \cap J_m|}{q}$ grows too steeply as a function of the overlap sizes. This rapid escalation risks inducing the well-known degenerate behavior associated with $k$-star and $k$-triangle statistics in exponential random graph models (ERGMs), potentially leading to unstable estimation processes or near-degenerate distributions \citep{hunter2006inference}.
%%%da qui
%%%new

The inclusion of multiple subset repetition orders within a single model specification introduces further complications. Lower-order statistics are intrinsically nested within higher-order ones—since subset repetition of order $(p,q)$ is a function of all overlaps of size at least $p$ and $q$, respectively—and this redundancy results in severe multicollinearity. Such interdependence poses significant challenges for both the numerical estimation and the substantive interpretation of individual effects.

The Geometrically Weighted Subset Repetition (GWSR) addresses this issue (Problem 2) by collapsing the contributions of all possible subset orders into a single, parsimonious statistic. For a two-mode hyperevent $(I, J)$, it is defined as follows: 
\begin{eqnarray*}
  \label{eq:gwsubrep_dir}
  gwsr^{(\kappa,\lambda)}(t,I,J)&=&\frac{\exp(\kappa+\lambda)}{|I|\cdot|J|}\sum_{t_m<t}\left\{1-\left(1-\frac{1}{\exp(\kappa)}\right)^{|I\cap I_m|}\right\}\cdot|I\cap I_m|\\
  &&\times
  \left\{1-\left(1-\frac{1}{\exp(\lambda)}\right)^{|J\cap J_m|}\right\}\cdot|J\cap J_m|
  \enspace.
\end{eqnarray*}

In this formulation, the inclusion of the scaling term $1/(|I||J|)$ ensures that the statistic is normalized relative to the current hyperevent size (Problem 3), making it comparable across events of varying cardinalities. Furthermore, replacing the binomial coefficients with convergent geometric functions ensures numerical stability. By bounding the growth of the statistic as overlaps increase, the GWSR effectively prevents the model degeneracy (Problem 1) typically associated with the steep escalation of standard subset repetition counts.

%do we want to add pseudocode???
\begin{figure}
    \centering
    \includegraphics[width=0.8\linewidth]{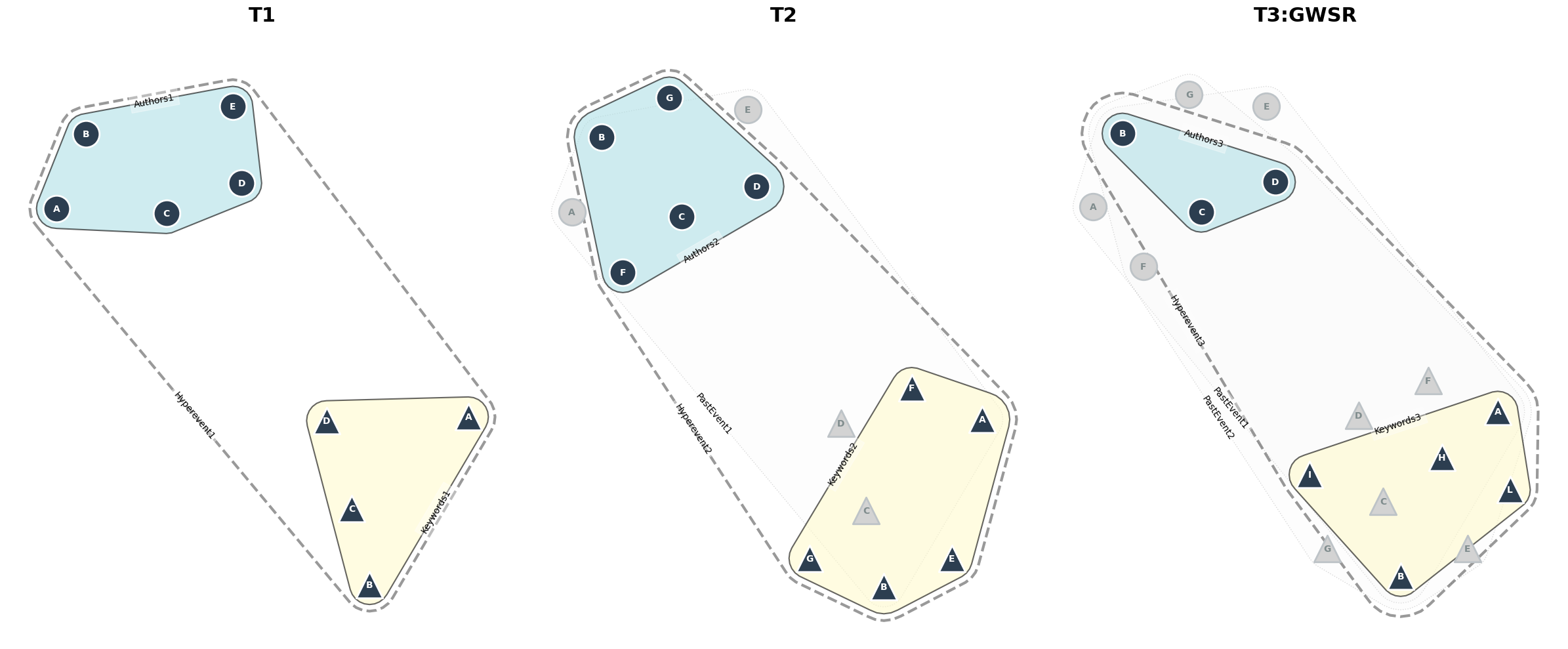}
    \caption{Sequence of three hyperevents $(t_1,I_1),(t_2,I_2),(t_3,I_3)$ composed by authors ($I$) and keywords ($K$). 
At each time point, the hyperevent is highlighted together with its subsets of authors (`Authors$i$) and keywords (`Keywords$i$), while past hyperevents (`PastEvent$j$, $j<i$) are dimmed in the background.
Darker nodes represent participants in the event, whereas grey nodes represent participants in past events. 
This image shows how overlapping subsets $(I \cap I_m)$ and $(K \cap K_m)$ overlap over time.
For example, between $t_1$ and $t_2$, the past hyperevent $I_m=\{A,B,C,D,E\}$ (PastEvent1) and the new hyperevent $I=\{B,C,D,F,G\}$ (Hyperevent2) share the intersection $I\cap I_m=\{B,C,D\}$, whose size directly affects the GWSR weighting.}
    \label{fig:placeholder}
\end{figure}

\subsubsection{Role of the scaling parameter $\kappa$ and $\lambda$}
The parameters $\kappa$ and $\lambda$ control the sensitivity of the statistic to overlapping hyperedges in the source and target sets, respectively. Specifically, $\kappa$ regulates the weighting of repeated source nodes $(I)$, while $\lambda$ regulates the target nodes $(J)$.

In general, the factor $\left\{1-\left(1-\frac{1}{\exp(\kappa)}\right)^{|I\cap I_m|}\right\}$ converges to one (from below) when the size of the intersection $|I_m\cap I|$ tends to infinity.

%% daqui the higher the value of $\kappa\geq 0$, the higher is the relative factor scaling the contribution of hyperedges $I_m$ that have a large overlap $|I_m\cap I|$. 

If $\kappa=0$ (or $\lambda=0$), geometrically-weighted subset repetition is identical to subset repetition of order one. 
In this case, only the past activity of individual nodes is counted, regardless of whether they co-participated with others in past events. 

If $\kappa>0$ (or $\lambda>0$), convergence is computationally fast for small values (meaning that small and large overlaps are similarly weighted). It measures repeated activity of individuals and, to some extent, repeated co-participation -- but it does not matter much whether this repeated co-attendance is with few or many others.

For large values, convergence is slow (i.e., large overlap is weighted much more than small overlap). Thus, if one is interested in repeated co-participation of rather large subsets, a higher value of $\kappa$ or $\lambda$ should be chosen. This ensures numerical stability, preventing model degeneracy that would arise from subset repetition and the steep increase in binomial coefficients (Problem 1), i.e., the increasing weight of past events as binomial coefficients grow.

Furthermore, this implementation allows the exclusion of either the source or the target node set. This is controlled by setting the corresponding parameter, $\kappa$ or $\lambda$, to a negative value (e.g., $-1.0$). In such cases, the negative value acts as a logical trigger: the geometric term associated with that dimension is removed from the product in Equation \ref{eq:gwsubrep_dir}, and the normalization factor is adjusted to reflect the single active dimension.

For instance, if $\lambda < 0$, the statistic ignores target set repetition and collapses to a unidimensional GWSR focused solely on the source nodes $I$:
\begin{equation}
  \label{eq:gwsubrep_collapsed}
  gwsr^{(\kappa, \text{neg})}(t,I,J) = \frac{\exp(\kappa)}{|I|} \sum_{t_m < t} \left\{ 1 - \left( 1 - \frac{1}{\exp(\kappa)} \right)^{|I \cap I_m|} \right\} \cdot |I \cap I_m| \enspace.
\end{equation}

This flexibility is crucial for our tripartite analysis. By focusing on one dimension, we can isolate baseline effects, such as an author's individual productivity or a reference's overall citation frequency, independent of specific co-participation patterns. 

\subsection{Geometrically Weighted Subset
Repetition effects}
To capture the tripartite effects, we defined different GWSR statistics where the set of authors, keywords and references produces three bipartite hyperevents:
\begin{itemize}
    \item[1.] $(t, I, J)$: the authors $I$ cite the set of papers $J$
\item[2.] $(t, I, W)$: the authors $I$ use the set of keywords $W$
\item[3.] $(t, W, J)$: a paper with a set of keywords $W$ cites the set of papers $J$
\end{itemize}

When modeling relational hyperevents of the form $(t, X, Y)$, where $(X, Y)$ represents one of the possible interactions between authors $(I)$; keywords $(W)$ and cited papers $(J)$ we distinguish three types of GWSR based on the configuration of the parameter $\kappa$ and $\lambda$:
\begin{itemize}
    \item[1.] GWSR only within $X$ (e. g., repeated coauthoring), where parameter $\lambda$ is negative, e. g., $(5,-1)$
\item[2.] GWSR only within $Y$ (e. g., repeated co-citations), this means that the parameter $\kappa$ is negative, e. g., $(-1, 5)$
\item[3.] GWSR within the two-mode hyperedge $(X, Y)$ (e. g., same authors repeatedly cite the same references), this means that both $\kappa$ and $\lambda$ are non-negative, e. g., (5, 5)
\end{itemize}

We defined different GWSR statistics, each capturing a specific dimension of scientific behavior. Within the model for $(I, J)$, that is, authors-references, we defined three statistics: 1) \textit{sub.rep.author} that reflects the author propensity to publish in future events; 2) \textit{sub.rep.reference} captures the tendency of a set of papers to be repeatedly co-cited by any group; 3)\textit{sub.rep.author.reference} identifies the author's propensity to repeatedly cite the same set of references. In the model for $(I, K)$, that is, authors-keywords: 1) \textit{sub.rep.keyword} identifies the persistence of a set of keyword over time; 2)\textit{sub.rep.author.keyword} captures the tendency of a set of authors to use the same set of keywords. Finally, in the model for $(K, J)$, that is, keywords-references, where we specified the \textit{sub.rep.keyword.reference} that indicates the propensity for associating the same keywords with the same set of references.

Using GWSR and combining it with closure, we can test different network assumptions and assess whether it stabilizes due to social ties, thematic similarity, or co-citational behavior, or by making other assumptions about network dynamics.

\section{Data and research questions}

The Relational Hyperevent Models can effectively capture the complex dependencies 
arising in tripartite networks of authors, references, and keywords. In particular, the novelty of this approach lies in the possibility of considering every event not as made by a single user, but in capturing the interactions among users, keywords, and references that constitute an event. We expect that models including closure effects involving different node sets (e.g., Author-Reference-Keyword, Keyword-Author-Reference) 
provide explanatory power beyond what is captured by closure effects 
(e.g., Author-Author-Author, Keyword-Keyword-Keyword) by taking into account both the co-evolution of the three entities and their internal evolution. 

Concretely, we address the following empirical research questions.
\begin{enumerate}
    \item[\textbf{RQ1}] Do we find evidence for effects specified via a third type of nodes (e.\,g., keywords) on top of all effects that can be specified using only two types of nodes (e.\,g., authors and references)? Does scientific collective production occur because of a complex interplay between three important determinants, that are social (i.e., common co-authors), previous research (i.e., common references) and common or complementary expertise (i.e., partially common topics)?
    \item[\textbf{RQ2}] Are findings related to one or two types of nodes affected, quantitatively and qualitatively, if we ignore any effects specified via a third type of nodes? Does the omission of one of the three model types affect the understanding of the whole system and consequently the understanding of the remaining node types?
    \item[\textbf{RQ3}] How does the model fit of the ``full model'' (specified with all possible interactions among the three types of nodes: 24 effects) relate to the model fit of sub-models obtained by dropping one type of nodes (i.\,e., specified with all possible interactions among the two remaining types of nodes: nine effects) and which of the three types of nodes (authors, references, and keywords) makes the largest contributions to the model fit on top of all effects specified with the two other types of nodes? By measuring model fit, can we determine the most important node set in the scientific network's knowledge production process? Which of the three node types best describes the organization of a scientific collaboration network?
    
    \item[\textbf{RQ4}] What is the relative contribution to the model fit of each of the 24 effects, on top of the model specified with the remaining 23 effects?
Which relational mechanism constitutes the most influential interaction pattern for modeling a scientific collaboration network?
\end{enumerate}

\subsection{Case study: the Italian academic statisticians community}

For the aim of this work, we studied the tripartite structure of authors-keywords-references of the Italian Academic Statisticians. Different studies have analyzed this community \citep{fabbrucci2025unveiling,bacci2023insights,destefanoetal_2023,de2013use} showing an emphasis on publishing in high-impact international journals, with distinct considerations across the five subsectors: 1) Statistics; 2) Statistics for Experimental and Technological Research; 3) Economic Statistics; 4) Demography; 5) Social Statistics.

The data were collected from the Italian Ministry of Education (MUR, Ministero dell'Istruzione e della Ricerca) \citep{MIUR}. For each year from 2014 to 2024, we obtained information on individual researchers belonging to the target community.

The purpose of using this source, due to its accuracy and reliability, is to minimize potential issues such as misspellings and homonyms. However, manual controls were still needed in some cases.

Once these steps were completed, the Scopus IDs of each academic were retrieved to provide a unique identifier using a machine learning algorithm and fuzzy clustering, with affiliation, name, surname, and affiliation history used to match profiles. This step is fundamental to the next phase of dataset creation, in which Scopus IDs are used to retrieve academic publications. Several checks were conducted to verify the correspondence between names, surnames, and their respective universities of affiliation.
The final dataset was created using Python and the Pybliometrics library \citep{pybibliometrics}, including keywords, references, co-authors' Scopus IDs, and publication unique identifiers.

\begin{figure}
    \centering
    \includegraphics[width=0.8\linewidth]{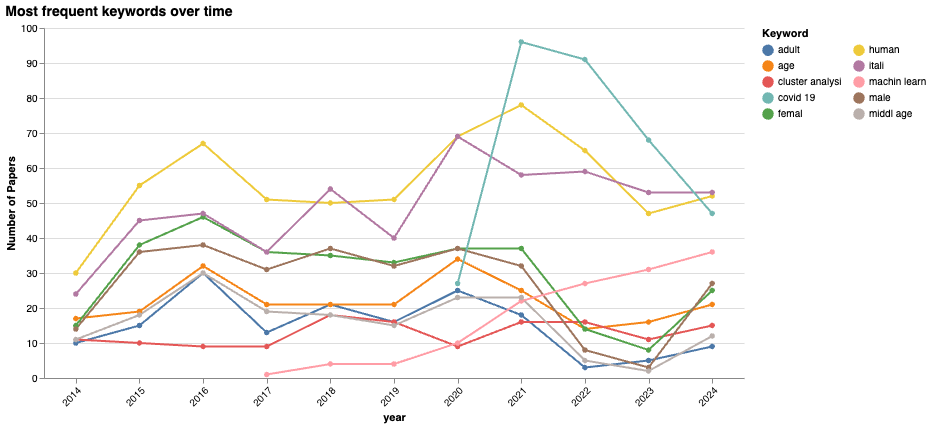}
    \caption{Most frequent keywords over time}
    \label{fig:monogram}
\end{figure}

\begin{figure}
    \centering
    \includegraphics[width=0.8\linewidth]{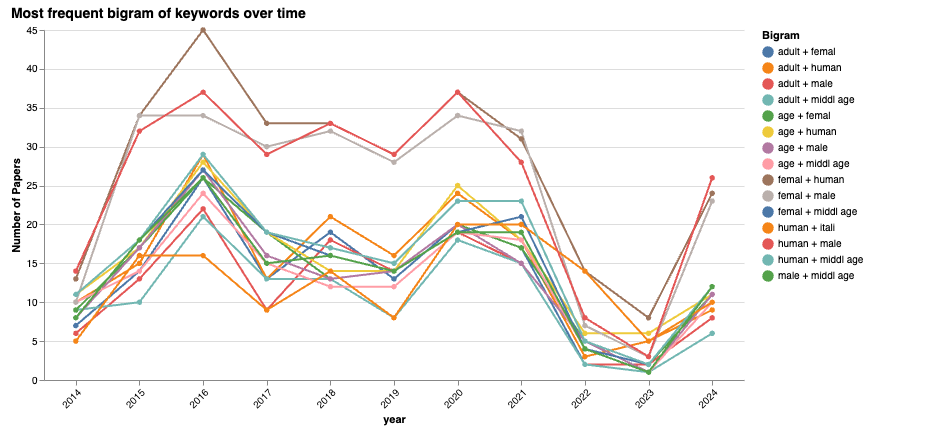}
    \caption{Most frequent bigram over time }
    \label{fig:bigram}
\end{figure}

\begin{table}[htbp]
\centering
\caption{Descriptive statistics of authors, keywords, and references per paper by year}
\label{tab:descriptive_stats_anno}
\begin{tabular}{ccccccccccccc}
\hline\hline
 &  & \multicolumn{3}{c}{Authors} & \multicolumn{4}{c}{Keywords} & \multicolumn{4}{c}{References} \\
\cmidrule(lr){3-5} \cmidrule(lr){6-9} \cmidrule(lr){10-13}
Year & N & Mean & Med. & Max & Mean & Med. & Max & \% & Mean & Med. & Max & \% \\
\hline
2014 & 1,013 & 6.10 & 3 & 400 & 4.47 & 5 & 22 & 86.4 & 36.62 & 30 & 1,211 & 97.7 \\
2015 & 1,076 & 5.69 & 3 & 875 & 4.83 & 5 & 26 & 87.4 & 36.03 & 32 & 269 & 97.3 \\
2016 & 1,133 & 6.30 & 3 & 394 & 5.14 & 5 & 28 & 89.8 & 37.78 & 33 & 299 & 97.1 \\
2017 & 1,153 & 5.36 & 3 & 384 & 5.07 & 5 & 27 & 90.1 & 39.63 & 36 & 361 & 96.9 \\
2018 & 1,189 & 6.47 & 3 & 372 & 4.95 & 5 & 29 & 90.6 & 42.21 & 38 & 525 & 97.1 \\
2019 & 1,251 & 9.69 & 4 & 1,363 & 4.99 & 5 & 21 & 91.8 & 45.09 & 38 & 606 & 97.4 \\
2020 & 1,388 & 13.62 & 4 & 3,099 & 4.94 & 5 & 29 & 89.4 & 46.20 & 40 & 987 & 96.8 \\
2021 & 1,610 & 6.71 & 4 & 289 & 4.99 & 5 & 37 & 90.1 & 47.72 & 42 & 543 & 97.3 \\
2022 & 1,511 & 6.33 & 4 & 284 & 4.80 & 5 & 27 & 90.3 & 49.55 & 44 & 2,139 & 96.5 \\
2023 & 1,427 & 7.61 & 4 & 507 & 4.72 & 5 & 17 & 89.8 & 49.94 & 45 & 375 & 97.2 \\
2024 & 1,581 & 6.43 & 4 & 343 & 5.08 & 5 & 24 & 91.1 & 52.51 & 47 & 738 & 96.9 \\
\hline
Overall & 14,332 & 7.38 & 3 & 3,099 & 4.91 & 5 & 37 & 89.8 & 44.73 & 39 & 2,139 & 97.1 \\
\hline\hline
\end{tabular}
\begin{tablenotes}
\small
\item \textit{Notes:} N indicates the number of papers per year. The table reports mean, median, and maximum values for authors, keywords, and references per paper, along with the percentage of papers containing keywords or references (\%). Overall statistics are reported in the last row.
\end{tablenotes}
\end{table}

The dataset includes 14,332 papers, with an average of 7.38 authors per paper and a median of 3, indicating a right-skewed distribution with a high proportion of highly collaborative papers.
The number of authors increases in 2019 (mean=9.69) and 2020 (mean=13.62), driven by large-scale collaborative studies during the COVID-19 pandemic; the latter paper includes 3099 authors, suggesting an international consortium. Keywords are stable (since the author usually indicates 4-5 keywords per paper) during the study period. The mean number of keywords per paper is 4.91, while the median is five across all years. Coverage is consistently high, with 89.8\% of papers overall containing keywords, ranging from 86.4\% to 91.8\% across years. As shown in Figure \ref{fig:monogram},  the keyword "human" is stable over time, while "COVID-19" emerged in 2020 and quickly became one of the most used terms by 2021. Furthermore, the bi-gram "Machine learning" shows gradual growth, reflecting the community's increasing interest in the topic over recent years. Demographic descriptors such as "adult," "male," "female," and "middle age" are relatively stable throughout the period, and the bi-gram "human + male" exhibits the highest frequency, as shown in Figure \ref{fig:bigram}.

Regarding the cited papers, the number of references per paper increases over time, with a mean that rises from 36.62 in 2014 to 52.51 in 2024, a 43\% increase over the decade. Overall, 97.1\% of the reference papers have been collected.

\section{Results}

Table~\ref{tab:four_models} reports estimated parameters, standard errors (in brackets), and model fit measured via the Akaike Information Criterion (AIC) of four models for publication events: the ``full model'' specified with all 24 effects dependent on any combination of the three types of nodes (authors, references, keyword) and the three sub-models obtained by dropping all effects dependent on one of the three types of nodes, respectively.

\begin{table}[ht]
\centering
\begin{tabular}{rlllll}
  \hline
 effect & Full model & Without AUTHORS & Without REFS & Without KEYS \\ 
  \hline
 sub.rep.aut & 0.64 (0.02)*** &  & 1.11 (0.02)*** & 0.62 (0.02)*** \\ 
 sub.rep.ref & -1.27 (0.04)*** & -1.29 (0.04)*** &  & -1.15 (0.03)*** \\ 
 sub.rep.key & 0.07 (0.02)*** & 0.04 (0.02) & 0.31 (0.02)*** &  \\ 
 sub.rep.aut.ref & -0.02 (0.02) &  &  & -0.17 (0.02)*** \\ 
 sub.rep.aut.key & 0.10 (0.02)*** &  & -0.12 (0.02)*** &  \\ 
 sub.rep.key.ref & -0.18 (0.02)*** & -0.19 (0.02)*** &  &  \\ 
 closure.ref.aut.ref & -0.22 (0.04)*** &  &  & 0.07 (0.02)*** \\ 
 closure.ref.ref.ref & 0.87 (0.02)*** & 1.41 (0.03)*** &  & 1.09 (0.02)*** \\ 
 closure.ref.key.ref & 0.63 (0.04)*** & 0.49 (0.03)*** &  &  \\ 
 closure.aut.aut.key & 0.08 (0.02)*** &  & -0.08 (0.02)*** &  \\ 
 closure.aut.key.key & -0.24 (0.03)*** &  & 0.08 (0.02)*** &  \\ 
 closure.aut.ref.key & -0.03 (0.04) &  &  &  \\ 
 closure.key.key.key & 0.17 (0.03)*** & 0.41 (0.02)*** & 0.08 (0.03)** &  \\ 
 closure.key.aut.key & 0.21 (0.04)*** &  & 0.20 (0.03)*** &  \\ 
 closure.key.aut.ref & 0.12 (0.03)*** &  &  &  \\ 
 closure.key.ref.ref & 0.62 (0.04)*** & 1.39 (0.03)*** &  &  \\ 
closure.key.key.ref & -0.48 (0.04)*** & -1.12 (0.03)*** &  &  \\ 
 closure.key.ref.key & -0.05 (0.03) & 0.11 (0.02)*** &  &  \\ 
 closure.aut.aut.ref & -0.59 (0.03)*** &  &  & -0.60 (0.02)*** \\ 
 closure.aut.ref.ref & 1.07 (0.04)*** &  &  & 0.91 (0.03)*** \\ 
 closure.aut.key.ref & -0.42 (0.04)*** &  &  &  \\ 
 closure.aut.aut.aut & -0.92 (0.03)*** &  & -0.78 (0.02)*** & -1.06 (0.03)*** \\ 
 closure.aut.ref.aut & 0.79 (0.03)*** &  &  & 0.84 (0.02)*** \\ 
 closure.aut.key.aut & -0.10 (0.03)*** &  & 0.66 (0.01)*** &  \\ 
   \hline
Number of events  &13915&           13915&        13915&        13915\\
Number of observations     &153065       &   153065       &153065       &153065\\
                    AIC     & 25080       &    34134        &34590  &      27229\\
              Delta AIC      &             &    9054         &9510   &      2149   \\
\hline
\end{tabular}
    \caption{Estimated parameters, standard errors (in brackets), and model fit measured via the Akaike Information Criterion (AIC) of four models for publication events: the ``full model'' specified with all 24 effects dependent on any combination of the three types of nodes (authors, references, keyword) and the three sub-models obtained by dropping all effects dependent on one of the three types of nodes, respectively.}
    \label{tab:four_models}
\end{table}

We find that all but a few parameters are significantly different from zero ($p<0.001$), even if we control for all 23 other effects operationalizing subset repetition and closure on any combination of the three types of nodes, positively answering RQ1. 

With respect to RQ2, we find that parameter values are often affected if we discard all effects related to any one of the three types of nodes. More relevant to substantive empirical research, we also find that, for some effects, the significance level and/or the sign of the parameters change. For example, the statistic \texttt{sub.rep.aut.key} has a significantly positive parameter in the full model, suggesting a tendency for the same authors to repeatedly use the same keywords in their publications. In contrast, the same effect is found to be significantly negative when we discard the impact of references in published papers. As another example, the statistic \texttt{closure.ref.aut.ref} has a significantly negative parameter in the full model, suggesting that two papers that have previously been cited by the same author (representing the ``third node'') are less likely to be co-cited in a future publication. In contrast, if we discard the impact of keywords, the same effect is found to be positive. In summary, with respect to RQ2, we find that for some effects, discarding the impact of any of the three node types can lead to substantively different conclusions.

Turning to the model fit (RQ3), we find that the full model has the lowest AIC of the four models shown in Table~\ref{tab:four_models}, indicating the best fit. With respect to the three sub-models, we find that discarding the references of publications (Model ``Without REFS'') has the most deteriorating effect on the model fit, closely followed by discarding the authors of publications. Discarding the keywords of publications also decreases model fit -- but the difference is much smaller than for the other two types of nodes. This result indicates that, in the considered dataset, we cannot ignore references, as they constitute the most influential node type in the overall organization of the collaboration network among Italian academic statisticians. 

In Table~\ref{tab:effects contributions} we report the contributions to the log-likelihood, and to the AIC, that each of the 24 effects makes on top of all 23 other effects, related to RQ4. (Formally, this is the difference of the log-likelihood, or AIC, of the full model minus the log-likelihood, or AIC, of the sub-model obtained by removing the effect listed in the first column.) We find that the four strongest effects use only the references or only the authors of publications. Concretely, the strongest effect (\texttt{sub.rep.ref}) is the effect to (co-)cite a set of papers with a higher probability if this set of papers has been more often (co-)cited before. Since the ``set of papers'' can also just comprise a single paper, this includes the well-established Matthew-effect or preferential attachment effect in citations; in addition, the effect also captures repeated co-citations to sets of two or more papers. 

The second strongest effect (\texttt{closure.ref.ref.ref}) could alternatively be denoted as ``co-citation closure''. It is the tendency to co-cite two papers $j$ and $j'$ that have both been co-cited with a common ``third paper'' before (possibly in two separate citing papers). 

The third strongest effect (\texttt{closure.aut.aut.aut}), alternatively denoted as ``co-authoring closure'', is the tendency that two scientists who share a common co-author become co-authors themselves. 

We also find that two effects contribute negatively to the AIC, implying that removing them from the full model would improve model fit. Consistent with that, these two effects have non-significant parameters in the full model in Table~\ref{tab:four_models}.

Overall, these effects combined describe an evolution of the collaboration network among Italian academic statisticians, mostly guided by the use of largely references (negative repetitions), indirectly connected (positive references closure), and a scientific trust mechanism among authors (positive authors closure).

\begin{table}[ht]
\centering
\begin{tabular}{rlrrrrrrr}
\hline
excluded effect & AIC & $\Delta$AIC & logLik & $\Delta$logLik \\ 
\hline
sub.rep.ref  & 27011.28 & 1933.51 & -13482.64 & -966.62  \\ 
closure.ref.ref.ref & 26155.80 & 1078.03 & -13054.90 & -538.88  \\ 
closure.aut.aut.aut & 26009.85 & 932.08 & -12981.92 & -465.91  \\ 
sub.rep.aut & 25939.04 & 861.27 & -12946.52 & -430.50  \\ 
closure.aut.ref.ref  & 25844.45 & 766.68 & -12899.22 & -383.21  \\ 
closure.aut.ref.aut & 25638.22 & 560.45 & -12796.11 & -280.09  \\ 
closure.aut.aut.ref  & 25574.32 & 496.55 & -12764.16 & -248.15  \\ 
closure.ref.key.ref & 25408.67 & 330.90 & -12681.33 & -165.32  \\ 
closure.key.ref.ref  & 25356.10 & 278.33 & -12655.05 & -139.04  \\ 
closure.key.key.ref  & 25238.45 & 160.68 & -12596.22 & -80.21  \\ 
closure.aut.key.ref  & 25191.19 & 113.42 & -12572.59 & -56.58  \\ 
sub.rep.key.ref  & 25174.85 & 97.08 & -12564.43 & -48.41  \\ 
closure.aut.key.key  & 25133.43 & 55.66 & -12543.72 & -27.70  \\ 
closure.ref.aut.ref  & 25126.76 & 48.99 & -12540.38 & -24.36  \\ 
closure.key.aut.key  & 25109.70 & 31.93 & -12531.85 & -15.83  \\ 
closure.key.key.key  & 25103.13 & 25.36 & -12528.56 & -12.55  \\ 
sub.rep.aut.key & 25101.86 & 24.09 & -12527.93 & -11.91  \\ 
closure.key.aut.ref & 25091.70 & 13.93 & -12522.85 & -6.84  \\ 
sub.rep.key  & 25090.12 & 12.35 & -12522.06 & -6.05  \\ 
closure.aut.key.aut & 25089.12 & 11.35 & -12521.56 & -5.55  \\ 
closure.aut.aut.key & 25088.45 & 10.68 & -12521.22 & -5.21  \\ 
closure.key.ref.key & 25081.17 & 3.40 & -12517.58 & -1.57  \\ 
closure.aut.ref.key  & 25079.02 & 1.25 & -12516.51 & -0.50  \\ 
sub.rep.aut.ref & 25078.90 & 1.13 & -12516.45 & -0.43  \\ 
\hline
\end{tabular}
\caption{Contribution of individual statistics to model fit (AIC). The baseline model is the "Full model" AIC=$25077.77$.}
\label{tab:effects contributions}
\end{table}

\subsection{Final discussion}
Hypergraphs are a generalization of graphs that can better represent complex network dynamics. In this work, we study the evolution of hypergraphs with three types of hyperedges and investigate the dynamics within and between them. We analyzed this using temporal dynamics, observing their evolution and obtaining a tripartite hyperevent. As a case study, we proposed a network of scientific productions involving authors, their citations, and keywords. By its nature, it results in a multi-partite process that does not involve only co-authoring but also other relational paths. While bipartite networks (e.g., author-references, author-keywords) provide a partial view that does not capture the co-evolution of different entities, the use of a tripartite structure (extendable to include additional entity types) and RHEM enable accounting for their contributions to network evolution.

This structure can be modeled as a tripartite hyperedge, which, through the application of the relational hypervent models, can be better investigated in its continuously evolving path. 
We argue that networks are systems of co-evolving entities, and that a collaboration network is an example in which their nature can be better understood by considering the combination of different entities and their persistence over time. In this context, the GWSR allows for weighting the overlap between new and past events, solving issues such as the combinatorial explosion of subset repetitions and the degeneracy of traditional counting statistics.  

Furthermore, we show that different combinations of closure among the three entities allow us to investigate distinct network behaviors, yielding fundamental statistics to study the coevolution of the three node types and to move from a bipartite to a tripartite structure.
Finally, the 24 effects allow us to study the different network behaviors and the impact of each of the three sets.
Our case study shows that removing any single node type decreases the $\Delta AIC$, indicating that scientific production cannot be reduced to a bipartite or one-mode relation, but rather to simultaneous interactions among different entities that define the evolution over time of the whole system.

\section{Preregistration statements}
This study was not preregistered.

\section{Declaration of conflict of interest}
The author(s) declared no potential conflicts of interest with respect to the research, authorship, and/or publication of this article.

\section{Data and code availability statements}
Data used in this study were obtained from Scopus (Elsevier) and cannot be made openly available due to licensing restrictions. Researchers with institutional access to Scopus may replicate the dataset using the search criteria described in Section 3.1. %The analysis code is provided as supplementary material with this submission. The eventnet configuration file has been provided as supplementary material during the submission process.

\bibliography{sn-bibliography}% common bib file
%% if required, the content of .bbl file can be included here once bbl is generated
\newpage

\section*{Appendix}
\begin{longtable}{m{3cm} m{4cm} m{6cm}}
    \caption{Summary of closure statistics implemented in the tripartite RHEM for authors ($I$), keywords ($K$), and references ($J$). } \label{tab:effects} \\
    \toprule
    \textbf{Structure} & \textbf{Closure} & \textbf{Explanation} \\
    \midrule
    \endfirsthead

    \multicolumn{3}{c}{{\bfseries \tablename\ \thetable{} -- Continued from previous page}} \\
    \toprule
    \textbf{Structure} & \textbf{Closure} & \textbf{Explanation} \\
    \midrule
    \endhead

    \bottomrule
    \multicolumn{3}{r}{{Continued on next page}} \\
    \endfoot
    \bottomrule
    \endlastfoot

    % --- SECTION 1: AUTHORS-REFERENCES (I, J) ---
    \noalign{\smallskip}
    \multicolumn{3}{l}{\cellcolor[gray]{0.9}\textbf{1. In the model for authors-references (I, J)}} \\
    \midrule
    \multicolumn{3}{l}{\textit{(a) authors $i_1$ and $i_2$ become coauthors; two path ($i_1, x, i_2$)}} \\
    \centering\includegraphics[width=2.5cm]{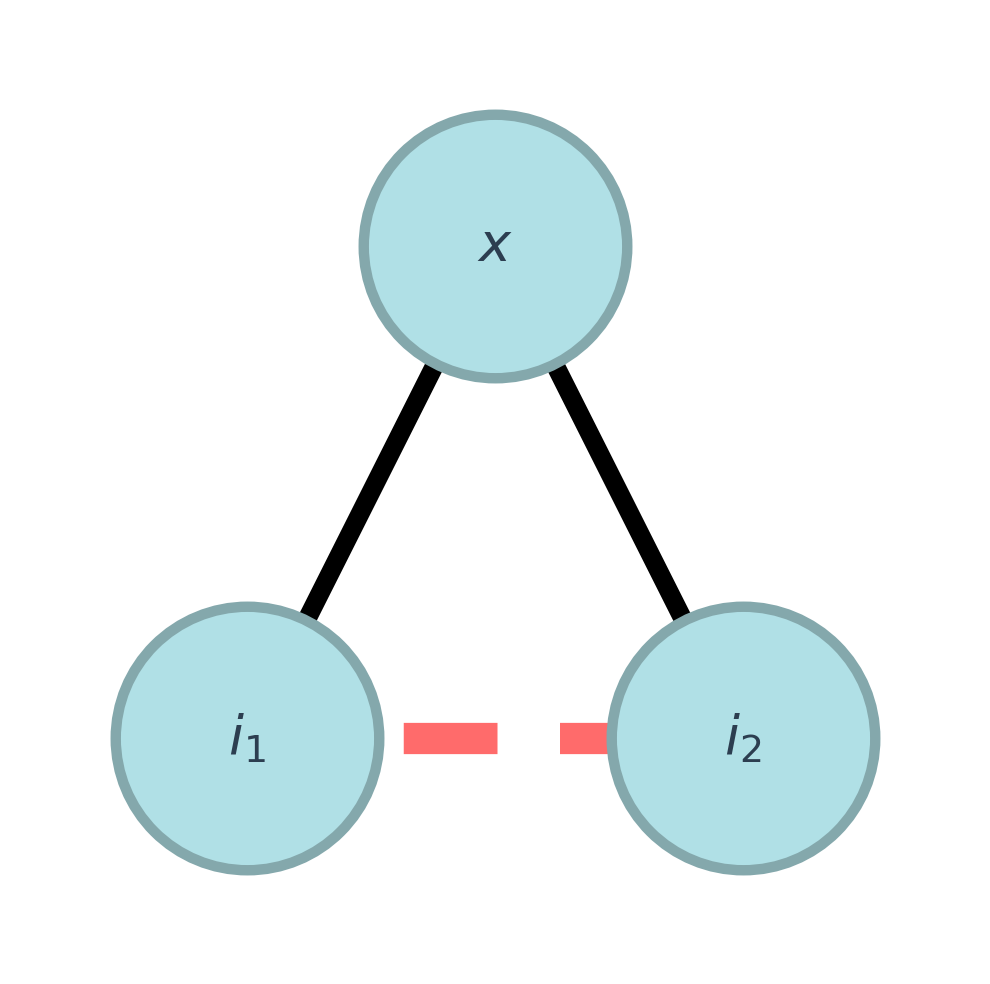} & Author-Author-Author & Authors tend to connect directly through another author \\
    \centering\includegraphics[width=2cm]{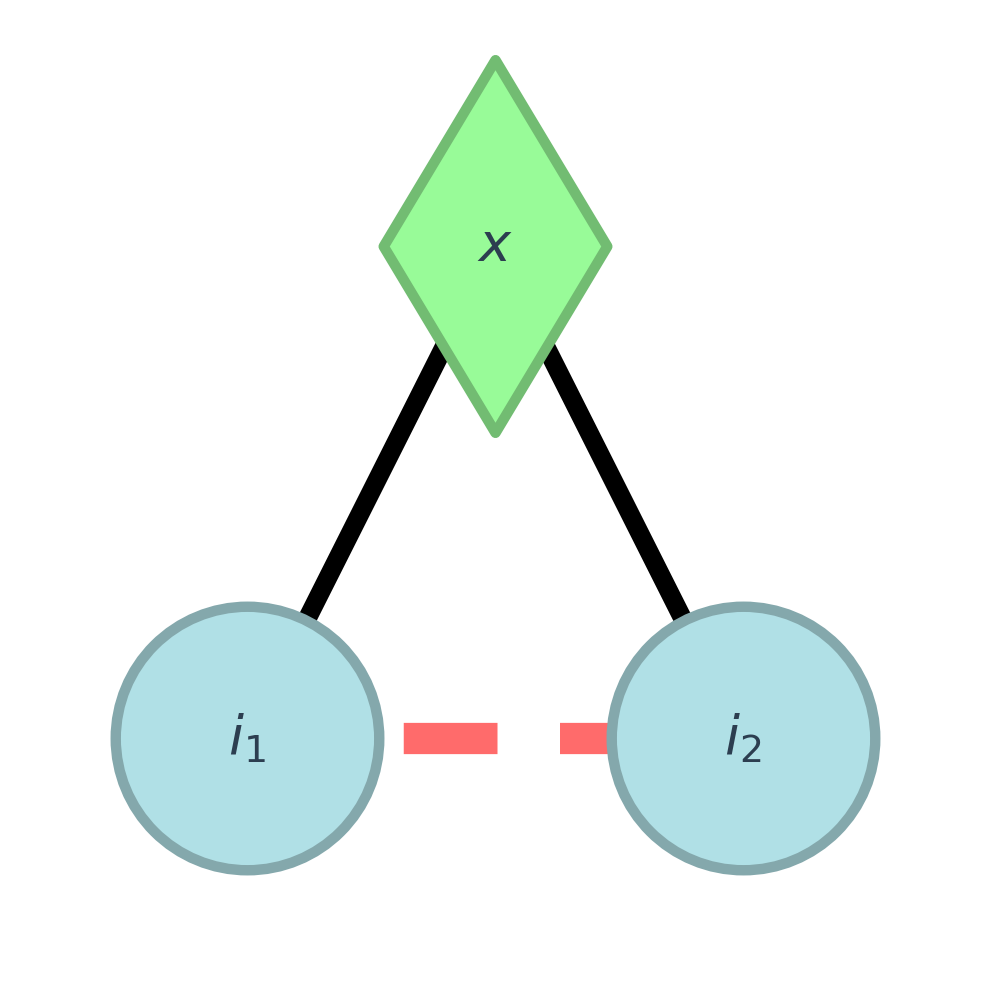} & Author-Reference-Author & Authors citing the same reference tend to be directly connected \\
    
    \noalign{\smallskip}
    \multicolumn{3}{l}{\textit{(b) papers $j_1$ and $j_2$ become cocited; two path ($j_1, x, j_2$)}} \\
    \centering\includegraphics[width=2.5cm]{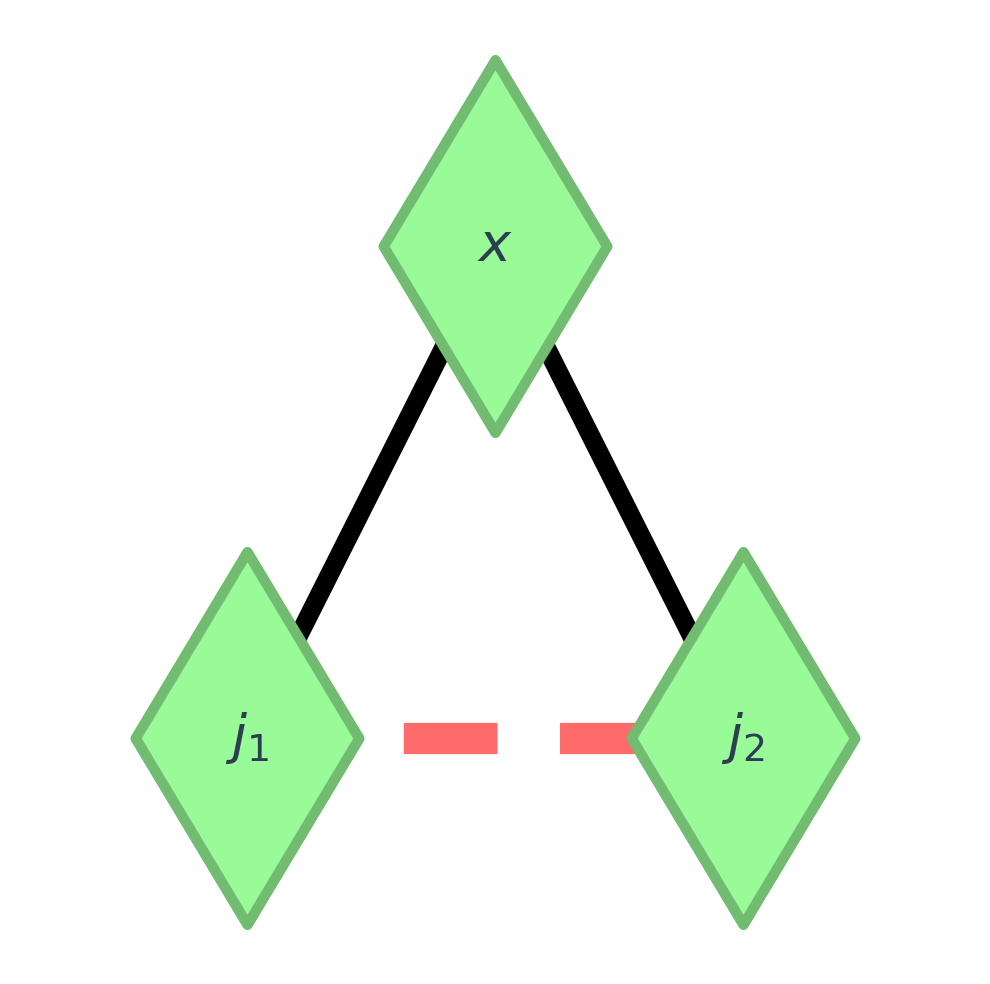} & Reference-Reference-Reference & Papers connected linked via common shared references tend to be directly linked \\
    
    \noalign{\smallskip}
    \multicolumn{3}{l}{\textit{(c) authors $i$ cite paper $j$; two path ($i, x, j$)}} \\
    \centering\includegraphics[width=2.5cm]{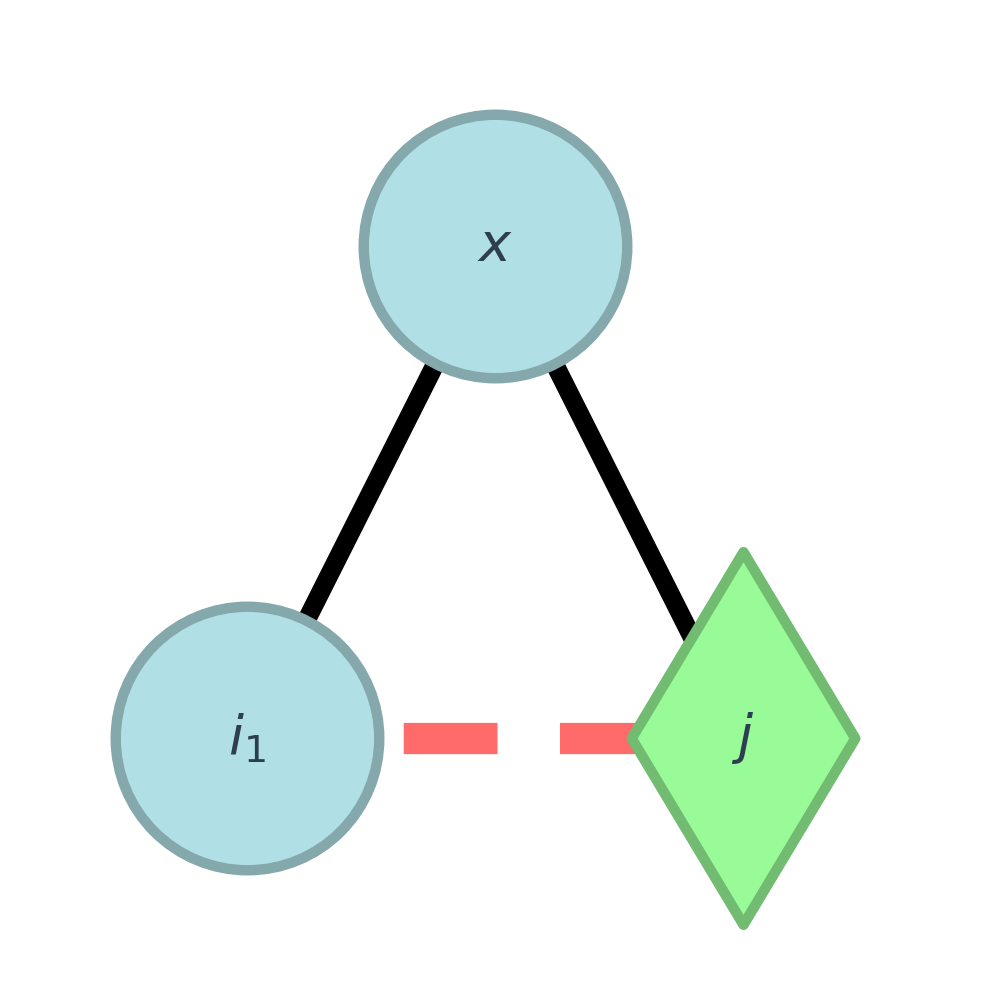} & Author-Author-Reference & An author tends to cite a paper that has been previously cited by their co-author \\
    \centering\includegraphics[width=2.5cm]{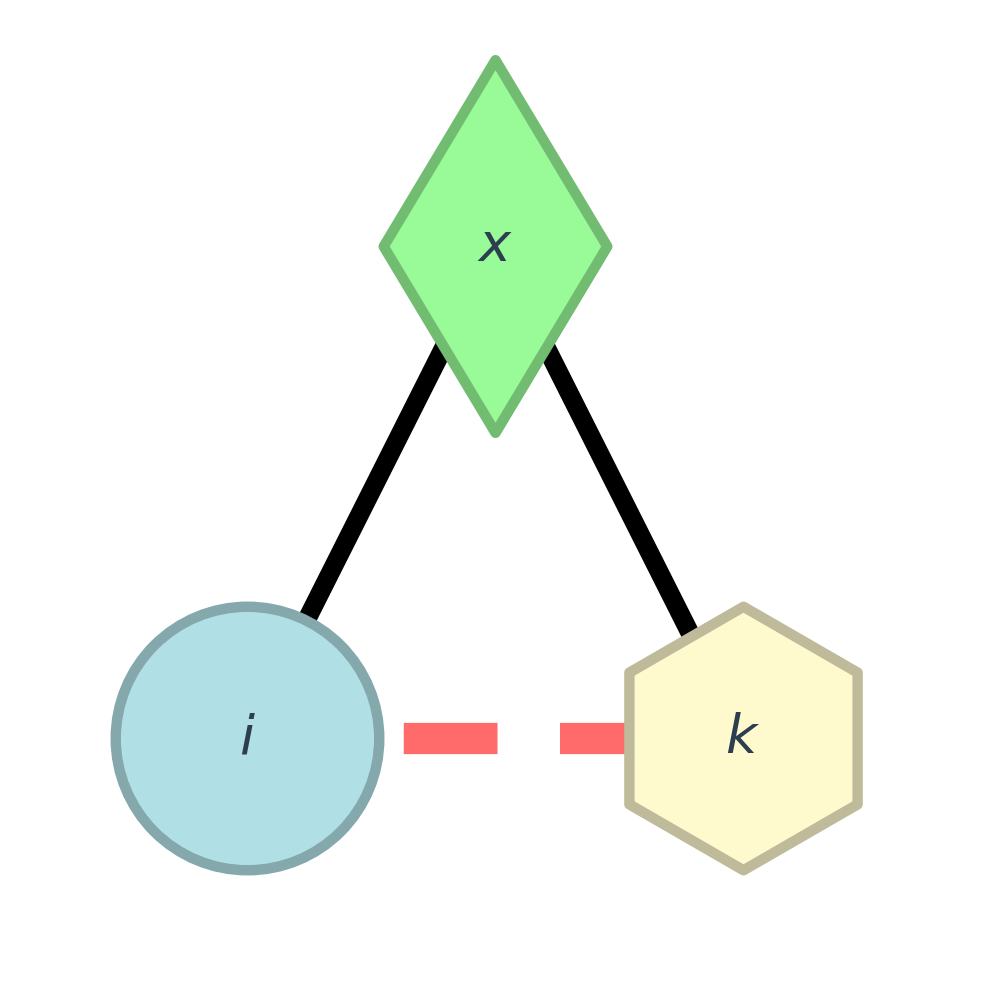} & Author-Reference-Keywords & An author citing papers labelled with a keyword tends to adopt that keyword \\

    \noalign{\bigskip}
    % --- SECTION 2: AUTHORS-KEYWORDS (I, K) ---
    \multicolumn{3}{l}{\cellcolor[gray]{0.9}\textbf{2. In the model for authors-keywords (I, K)}} \\

    \noalign{\smallskip}
    \multicolumn{3}{l}{\textit{(1) keywords $k_1$ and $k_2$ co-label a paper; two path ($k_1, x, k_2$)}} \\
    \centering\includegraphics[width=2.5cm]{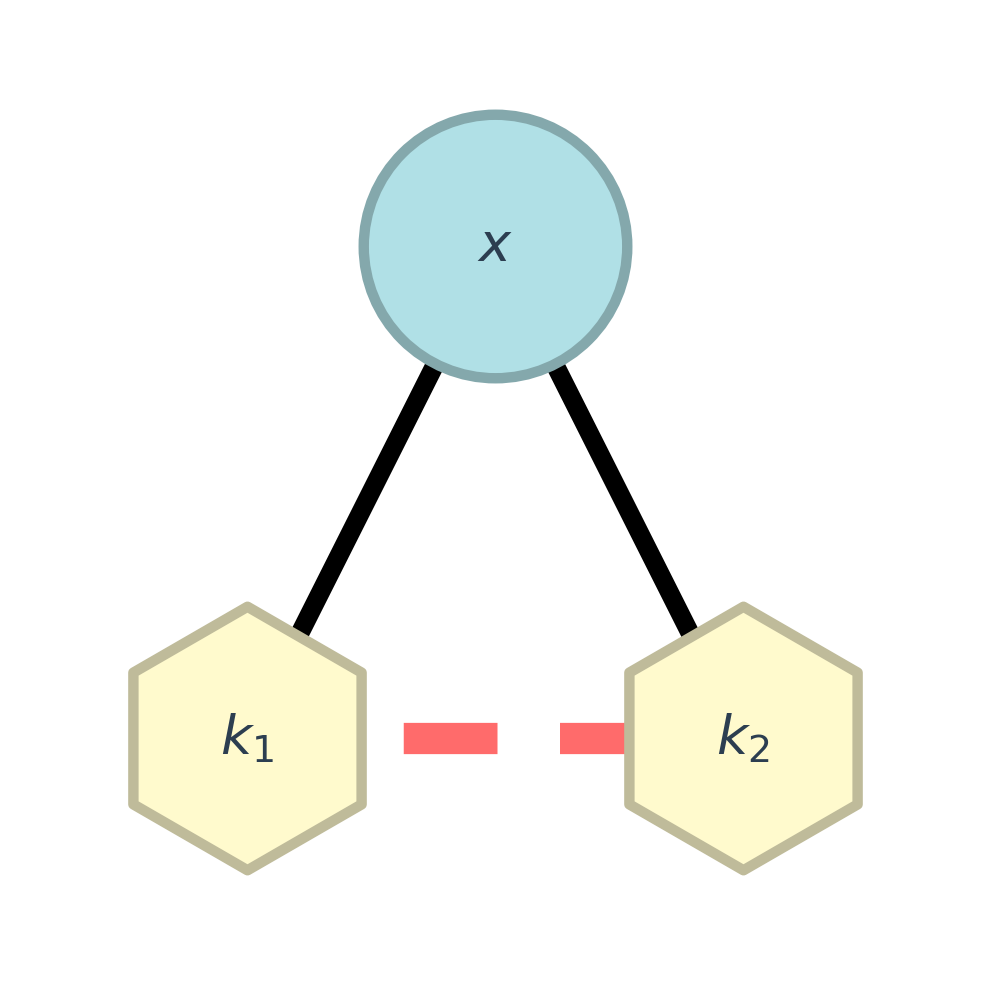} & Keywords-Authors-Keywords & Author uses similar keywords \\
    \centering\includegraphics[width=2.5cm]{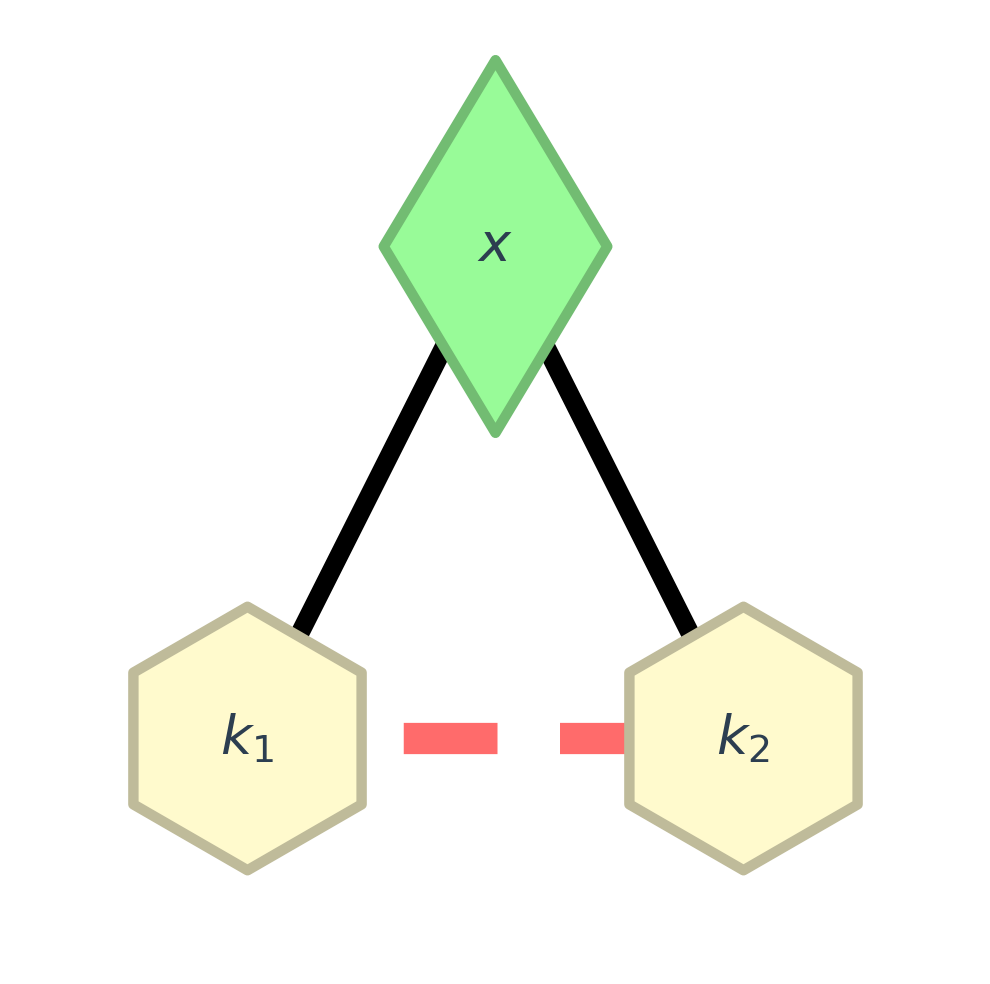} & Keyword-Reference-Keyword & Keywords labelling the same paper tend to co-occur %(semantic association via paper) 
    \\
    \centering\includegraphics[width=2.5cm]{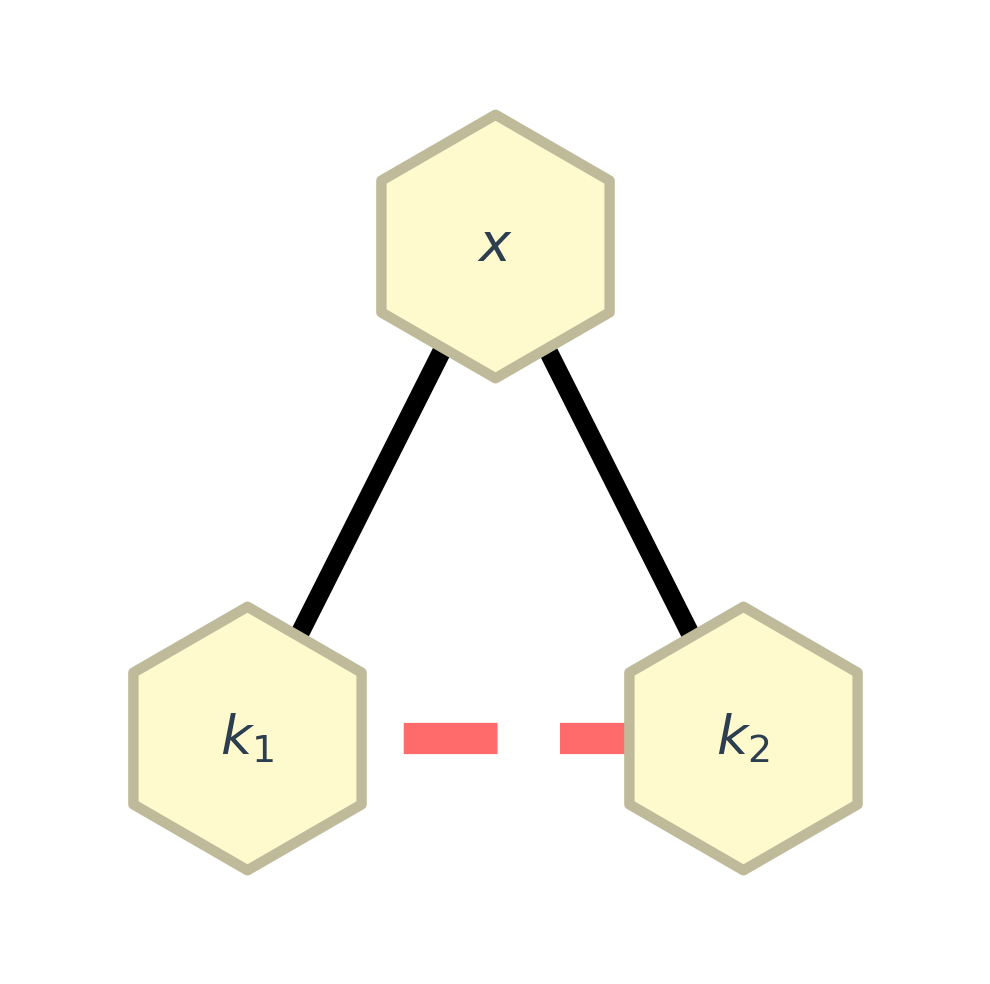} & Keyword-Keyword-Keyword & Keywords with a shared common keyword tend to connect directly. \\ %(Transitivity effect) 
    
    \noalign{\smallskip}
    \multicolumn{3}{l}{\textit{(b) authors $i$ use keyword $k$; two path ($i, x, k$)}} \\
    \centering\includegraphics[width=2.5cm]{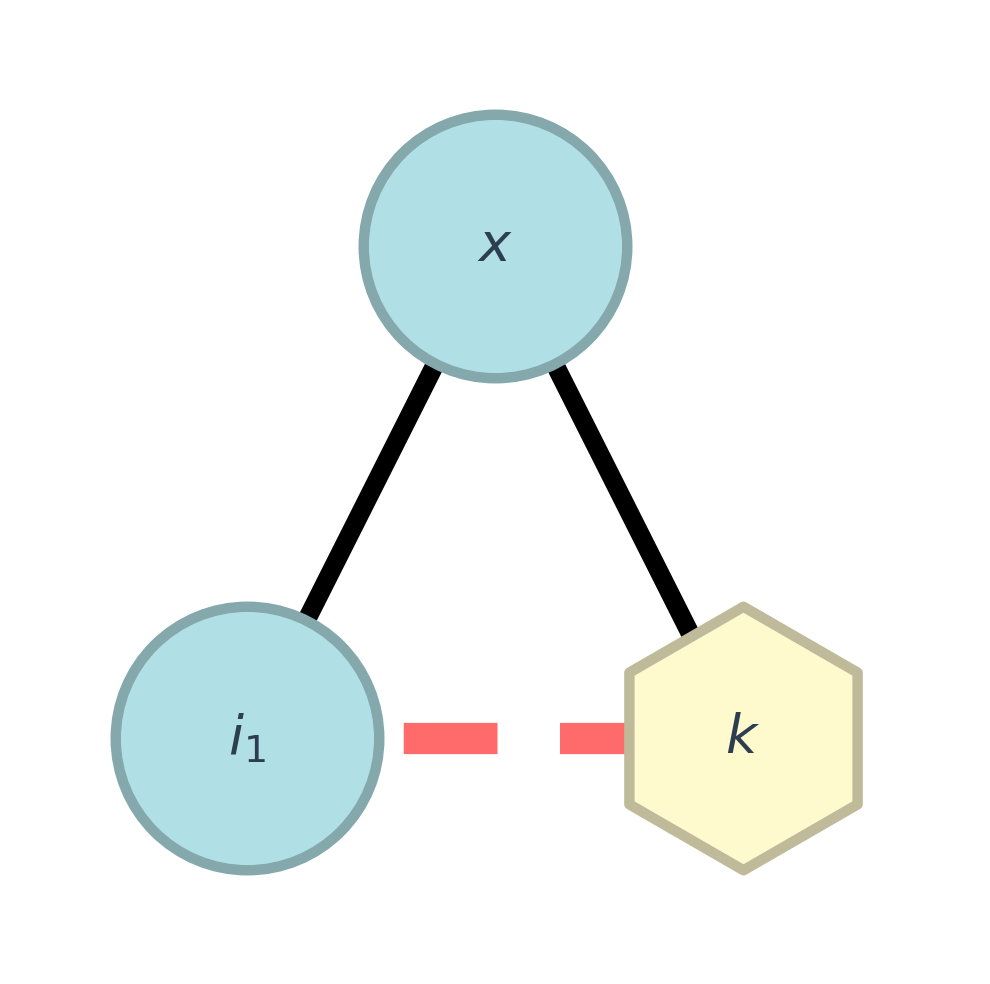} & Author-Author-Keywords & An author adopt the keywords used by other authors they are connected to \\
    \centering\includegraphics[width=2.5cm]{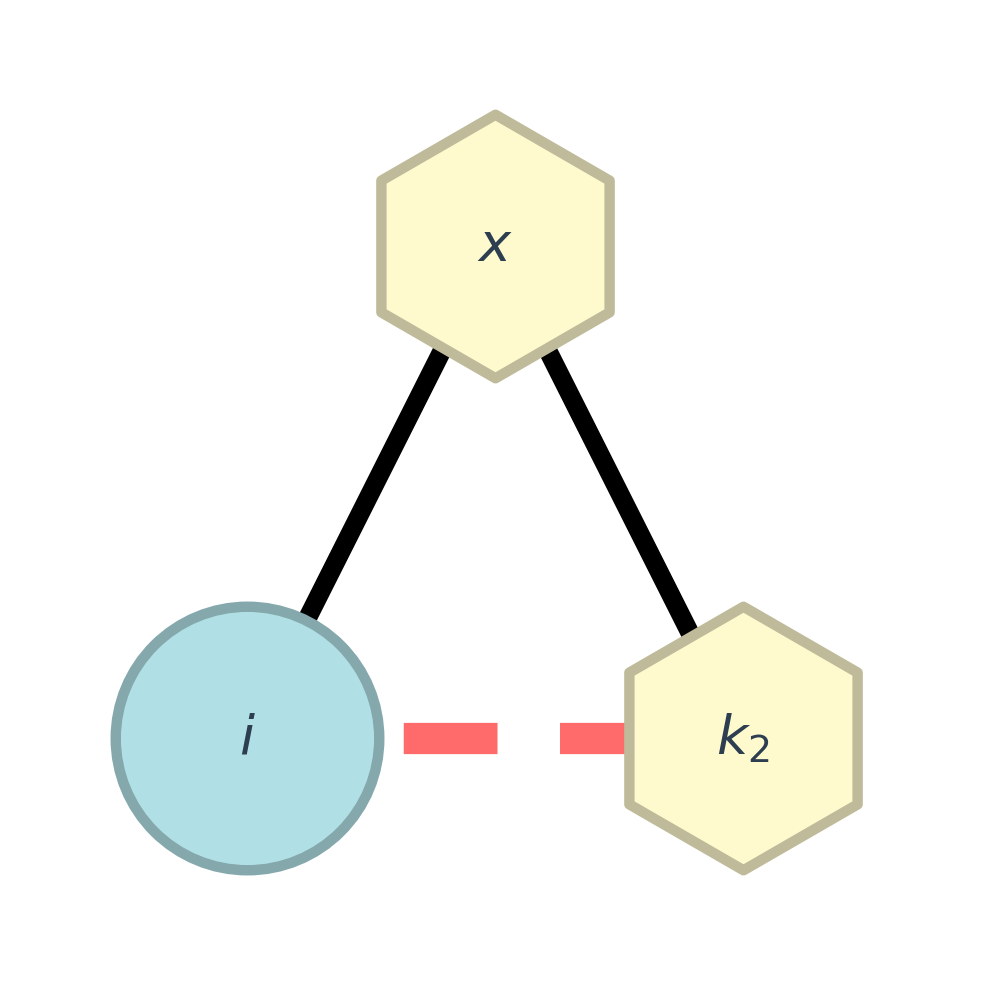} & Author-Keywords-Keywords & An author using keywords tends to adopt semantically similar keywords \\

    \noalign{\bigskip}
    % --- SECTION 3: KEYWORDS-REFERENCES (K, J) ---
    \multicolumn{3}{l}{\cellcolor[gray]{0.9}\textbf{3. In the model for keywords-references (K, J)}} \\
    \midrule
    \multicolumn{3}{l}{\textit{(a) keyword $k_1$ and $k_2$ co-label a paper; two path ($k_1, x, k_2$)}} \\
    \centering\includegraphics[width=2.5cm]{pictures/closure2/key.aut.key.png} & Keywords connected via common authors tend to co-appear  \\
    
    \noalign{\smallskip}
    \multicolumn{3}{l}{\textit{(b) papers $j_1$ and $j_2$ become cocited; two path ($j_1, x, j_2$)}} \\
    \centering\includegraphics[width=2.5cm]{pictures/closure2/ref.ref.ref.png} & Reference-Reference-Reference & Papers connected linked via common shared references tend to be directly linked \\
    
    \noalign{\smallskip}
    \multicolumn{3}{l}{\textit{(c) keyword $k$ labels a paper that cites $j$; two path ($k, x, j$)}} \\
    \centering\includegraphics[width=2.5cm]{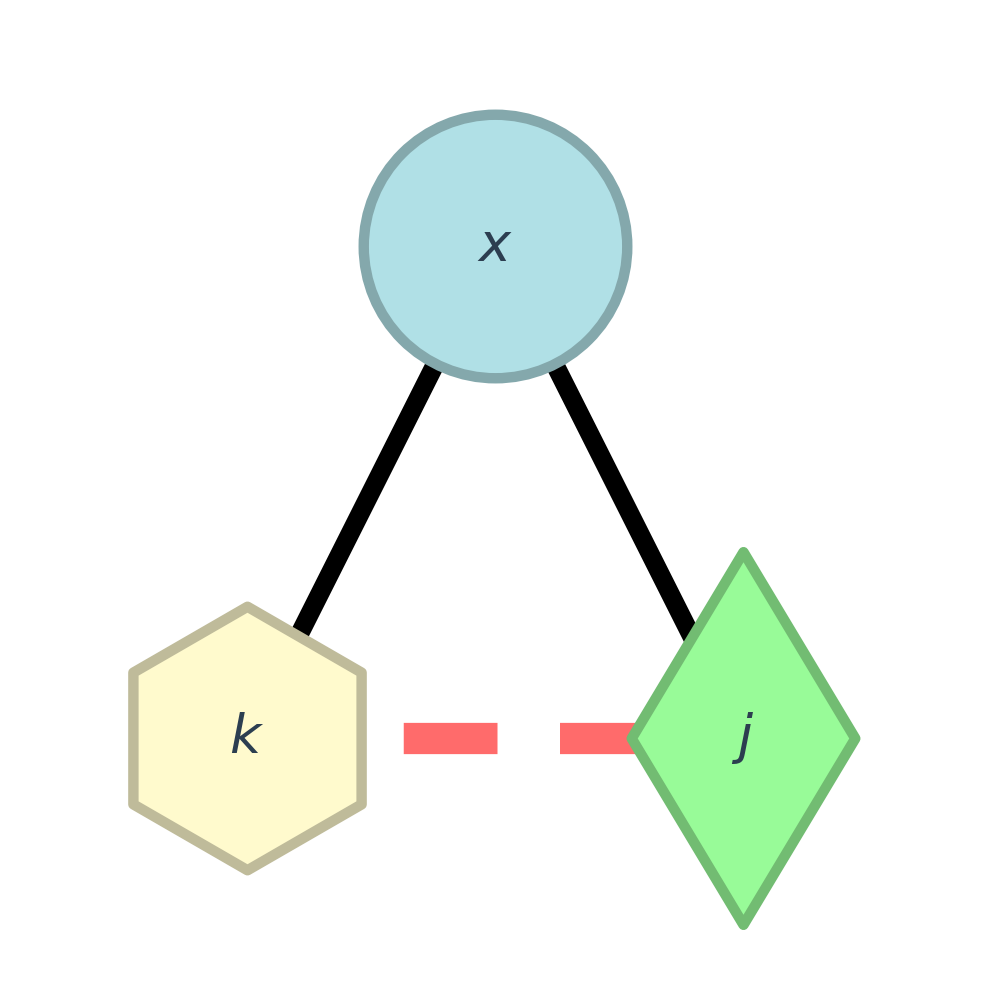} & Keyword-Author-Reference & If an author uses a keyword and cites a paper, that keyword tends to appear in the cited paper \\
    \centering\includegraphics[width=2.5cm]{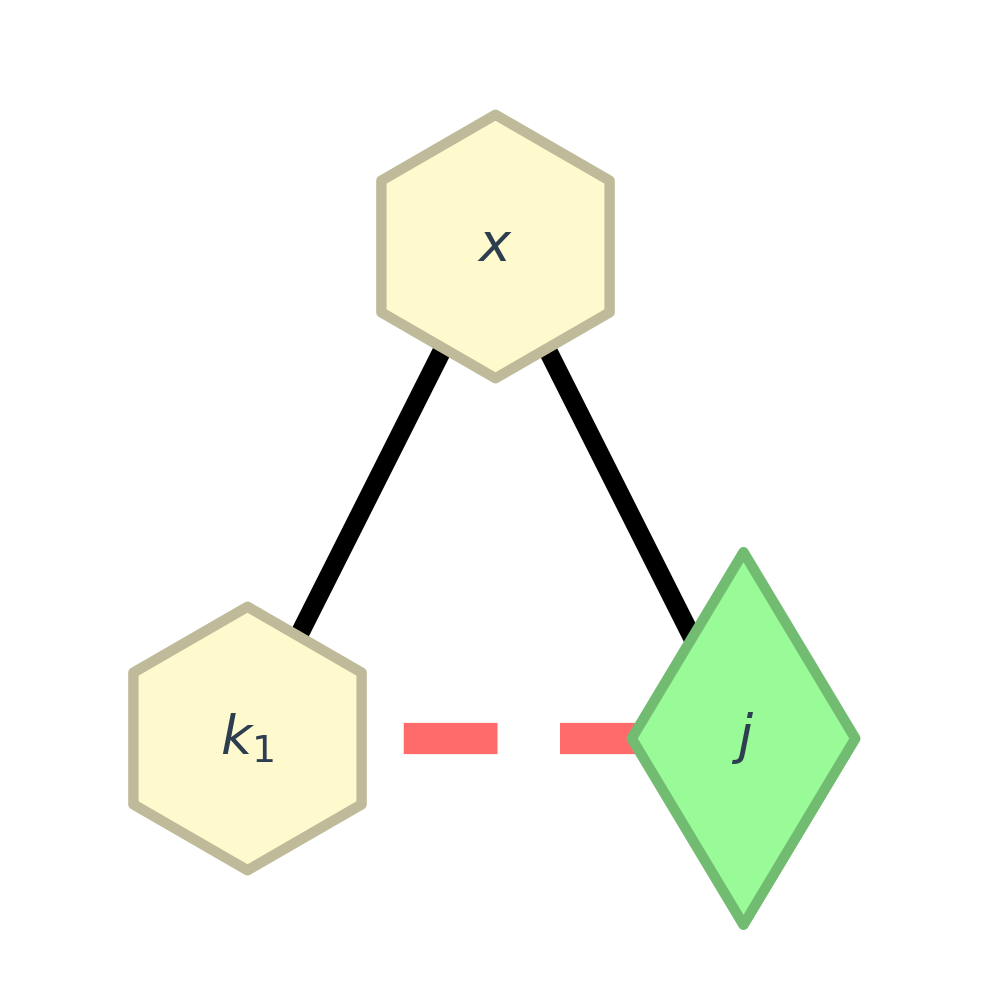} & Keyword-Keyword-Reference & Two connected keywords tend to label the same references. \\ %(semantic propagation) 
    \centering\includegraphics[width=2.5cm]{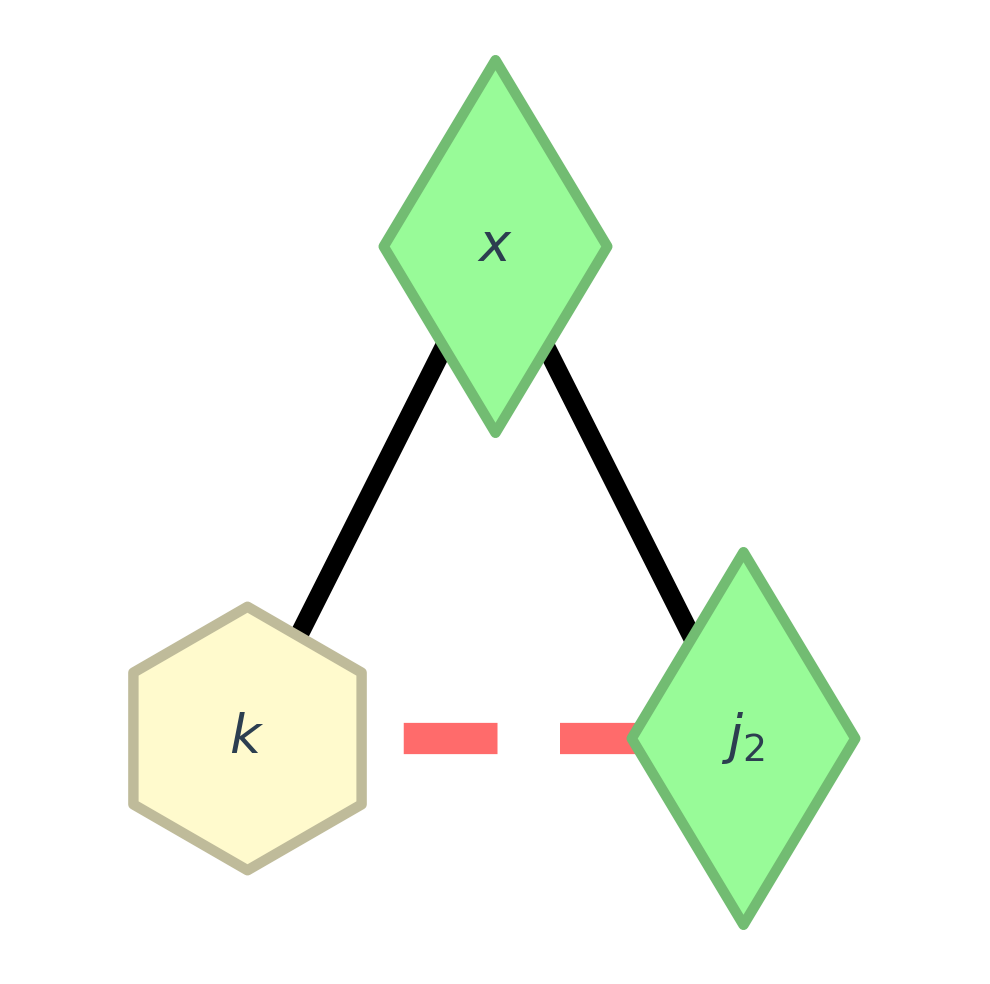} & Keyword-Reference-Reference & Keywords labelling a cited paper tends to be present in the cited paper references \\
    \centering\includegraphics[width=2.5cm]{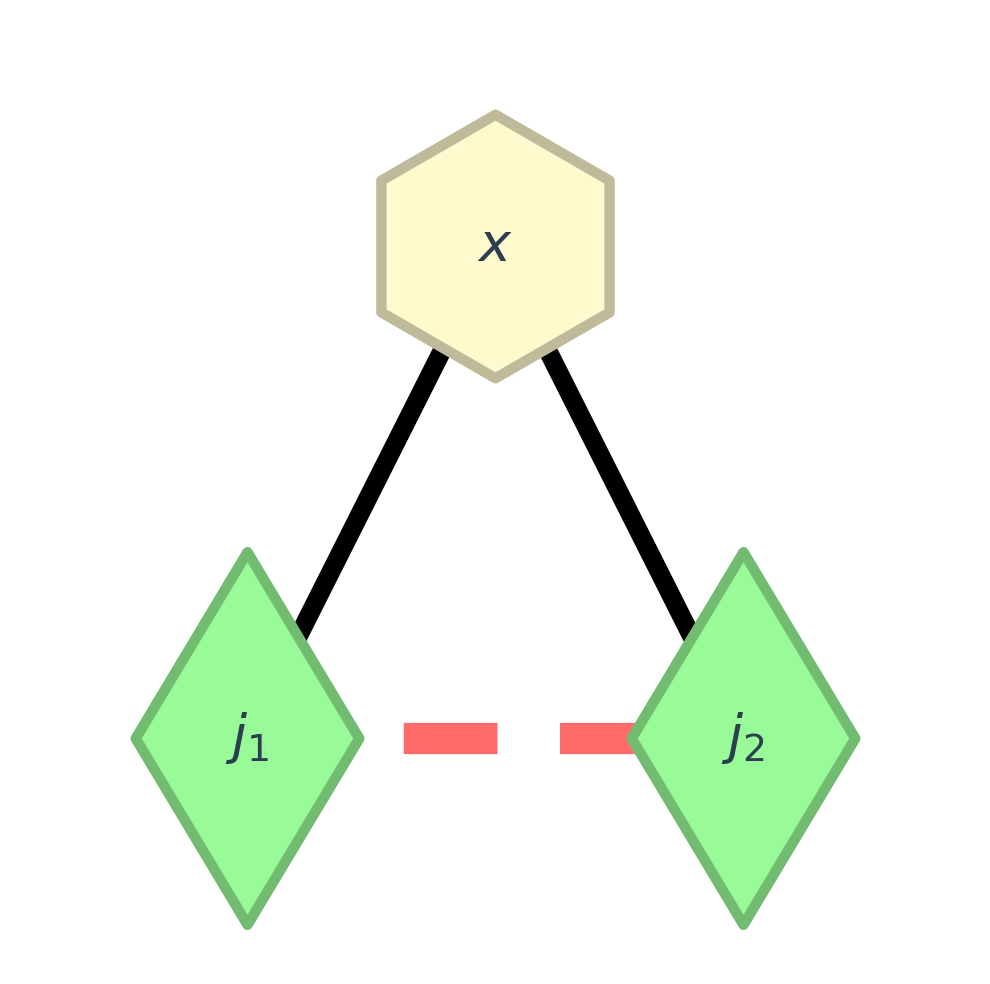} & Reference-Keyword-Reference & Papers with the same keyword are more likely to be co-cited or thematically linked \\
    \centering\includegraphics[width=2.5cm]{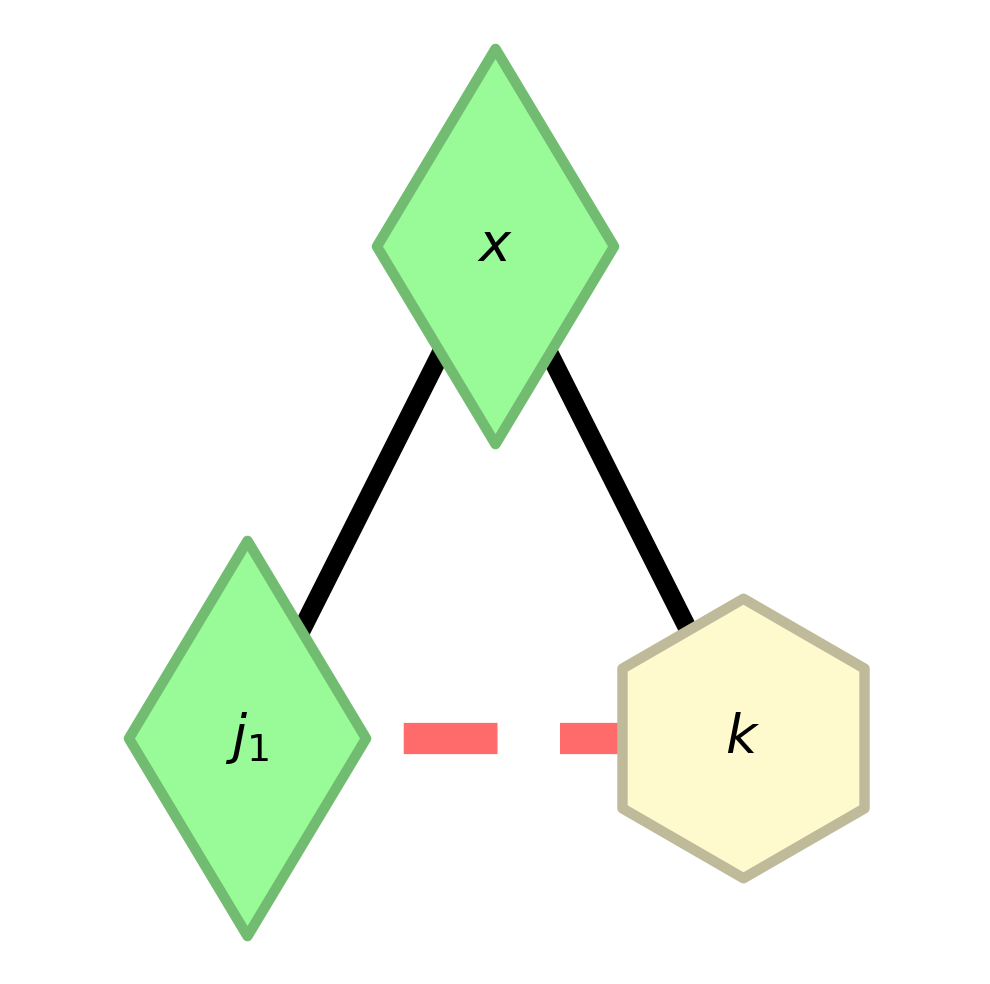} & Reference-Reference-Keyword & Keywords linked to a paper will extend to other papers bibliographically close \\

\end{longtable}

\end{document}